\documentclass[%
 reprint, superscriptaddress, 
 amsmath,amssymb, aps, 
 longbibliography
 ]{revtex4-1}
\usepackage{graphicx}
\usepackage{subfigure}
\usepackage{multirow}
\usepackage{comment}
\usepackage[usenames]{color}
\usepackage{dcolumn}
\usepackage{bm}
\usepackage[colorlinks=true, allcolors=blue]{hyperref}
\usepackage{amsmath}
\usepackage{mathtools}
\usepackage{xcolor}
\usepackage{dcolumn}
\usepackage{bm}
\usepackage[normalem]{ulem}

\usepackage{xr}
\graphicspath{ {./FiguresDef/} }

\begin{document}
\title{Consensus formation and relative stimulus perception in quality-sensitive, interdependent agent systems}

\author{David March-Pons}
\email{david.march@upc.edu}
\affiliation{Departament de Física, Universitat Politècnica de Catalunya, Campus Nord B4, 08034 Barcelona, Spain}

\author{Ezequiel E. Ferrero}
\affiliation{Departament de Física de la Matèria Condensada, Universitat de Barcelona, Martí i Franquès 1, 08028 Barcelona, Spain.}
\affiliation{Institute of Complex Systems (UBICS), Universitat de Barcelona, Barcelona, Spain
}
\affiliation{Instituto de Nanociencia y Nanotecnolog\'{\i}a, CNEA--CONICET, 
Centro At\'omico Bariloche, R8402AGP S. C. de Bariloche, R\'{\i}o Negro, Argentina.}

\author{M. Carmen Miguel}
\affiliation{Departament de Física de la Matèria Condensada, Universitat de Barcelona, Martí i Franquès 1, 08028 Barcelona, Spain.}
\affiliation{Institute of Complex Systems (UBICS), Universitat de Barcelona, Barcelona, Spain
}

\begin{abstract}

    We perform a comprehensive analysis of a collective decision-making model inspired by honeybee behavior. This model integrates individual exploration for option discovery and social interactions for information sharing, while also considering option qualities. Our assessment of the decision process outcome employs standard consensus metrics and investigates its correlation with convergence time, revealing common trade-offs between speed and accuracy. 
    Furthermore, we show the model's compliance with Weber's Law of relative stimulus perception, aligning with previous analysis of collective decision problems. Our study also identifies non-equilibrium critical behavior in specific limits of the model, where the highest values of consensus are achieved.
    This result highlights the intriguing relationship between optimal performance, critically, and the fluctuations caused by finite size effects, often seen in biological systems. Our findings are especially relevant for finite adaptive systems, as they provide insights into navigating decision-making scenarios with similar options more effectively.
\end{abstract}

\maketitle

\section{Introduction}

Collective decision making is an emergent, self-organized phenomenon by which a 
group of agents, each with its own decision mechanisms at the individual level, 
reaches an agreement on a certain task or topic~\cite{dyer2009, living_in_groups_krause,baronchelli_emergence_2018,bose2017}.
From humans taking part in elections, social mammal herds, flocking fish, foraging insect 
colonies or robot swarms, collective decision making is an ubiquitous process across all 
scales of biological and artificial complexity~\cite{dyer2009,living_in_groups_krause, collective_animal_beh_book, animal_signals_book, Leonard2024}. 
Thus, addressing this question requires interdisciplinary approaches integrating fields 
such as sociology, behavioral ecology, biology, physics, computer science and communication studies, 
among many others~\cite{dyer2009,sasaki2018}. 
Through empirical observations, field experiments, laboratory studies, 
and computational modeling, researchers strive to unravel the intricate 
dynamics of collective decision making and its significance.

Many efforts have been devoted in the modeling of opinion dynamics in 
order to understand how a collective makes a decision among a set of options. 
Initial attempts devised models that encompassed simple imitation rules 
controlling the build up of an agreement between equivalent options, 
such as the voter model~\cite{holley1975,castellano_statistical_2009} or 
the majority rule model~\cite{galam2002}. 
This models, though, have been regarded as too unrealistic in their basic 
formulation, and many more complex models have been proposed in order to 
capture a richer spectrum of social behaviors, such as partisanship, 
heterogeneity, personalized information or non linear interactions~\cite{galam2008, CastellanoPRE2009,redner_reality-inspired_2019,de_marzo_emergence_2020, bizyaeva_nonlinear_2023, RamirezPRE2024}.

In the realm of animal behavior, the existence of a remarkable diversity of 
animal signals has enthralled not only researchers studying collective motion 
and consensus formation but also non-experts, as these signals embody some 
of the most captivating facets of the natural world. 
Evolving to facilitate and enrich communication, these signals are conspicuous 
and compelling, even to casual observers. 
Animal signals prompt a myriad of scientific inquiries probing their functions, 
the information they convey, their production mechanisms, and the evolutionary 
processes driving them. These questions encapsulate the essence of comprehending 
animal communication and its significance in collective decision-making within 
the natural world.

In recent years, the collective behavior exhibited by social insects has 
become a focal point of interest within the scientific community. 
They conceive a paradigmatic example of systems formed by rather simple 
units exhibiting this kind of collective emergent phenomena.
In particular, the nest-site selection process in honeybee swarms, 
widely studied by biologists and 
ecologists~\cite{britton2002,Seeley2001NestsiteSI,seeley_stop_2012}, has laid 
the ground for a great number of collective decision making models 
inspired in this fascinating process~\cite{list_independence_2008,valentini2014,reina_desing_pattern,reina_model_2017,gray_multiagent_2018}, 
and for further exploration of these dynamics in robot swarms~\cite{reina_2016_decentdecision, valentini2016, talamali2021_less_more, marchpons2024_kilobots}.

In a bee-like house-hunting process, agents can adopt a wider spectrum of behaviors beyond 
the usual imitation rules. 
Initially, they must discover available options either through individual 
exploration or by communicating with peers using different signaling mechanisms. 
Additionally, they have the ability to abandon options and become uncommitted 
or neutral. 
Lastly, they may cross-inhibit other peers with different opinions. 
Importantly, in honeybee-inspired models, agents are able to estimate the 
qualities of options, which significantly impacts their subsequent signaling 
behavior~\cite{passino_swarm_2008}. 
Specifically, the better they estimate the quality to be, the longer they 
will advertise this option to their peers. 
This mechanism allows a positive feedback loop to regulate the group's 
opinion formation, discarding the worst option in favor of the more suitable ones. 
Ultimately, a collective agreement is reached through a combination of 
individual exploration of the environment to identify possible options, 
estimation of their quality, and information exchange among members of the group.

List, Estholtz and Seeley presented an agent-based model 
(in the following, the LES model) of the decision making 
process performed by honeybee swarms~\cite{list_independence_2008}, further 
developed analytically later by T. Galla~\cite{Galla_2010}, which focuses on 
these three features mentioned above.
The LES model is a simple model that introduces option qualities into the decision process. 
By balancing individual option exploration and social interactions, it effectively captures 
decision-making dynamics.
The main result highlighted in the original investigation is that when individuals possess sufficiently accurate estimates of the qualities and engage in a high degree of social 
interaction, the swarm tends to converge towards collective agreement, particularly 
when faced with various options of differing quality values.

In this paper, we delve deeper into studying the formation of consensus 
by inspecting the LES model. 
Our investigation focuses on the interplay between social feedback and 
individual exploration, which the model balances through a single parameter 
known as the ``interdependence'' parameter. 
The intensity of individual, independent exploration serves a dual purpose. 
On one hand, it is essential for the group to perceive all available options 
or to adapt in changing environments~\cite{talamali2021_less_more, aust_hidden_2022}. 
On the other hand, excessive individual exploration can introduce noise that hampers 
the effectiveness of the aforementioned positive feedback loop. 
Agents tend to overlook the actual state of their peers and adopt opinions based 
on the outcome of their own exploration. 
Consequently, stronger social interaction is required to balance this effect. 
Recent work highlights the capability of decentralized systems to effectively 
balance noise with social interactions under various scenarios~\cite{zakir_robot_2022,khaluf_interactionn_models_2018,rausch_coherent_2019}.

\begin{figure}
    \centering
    \includegraphics[width=.48\textwidth]{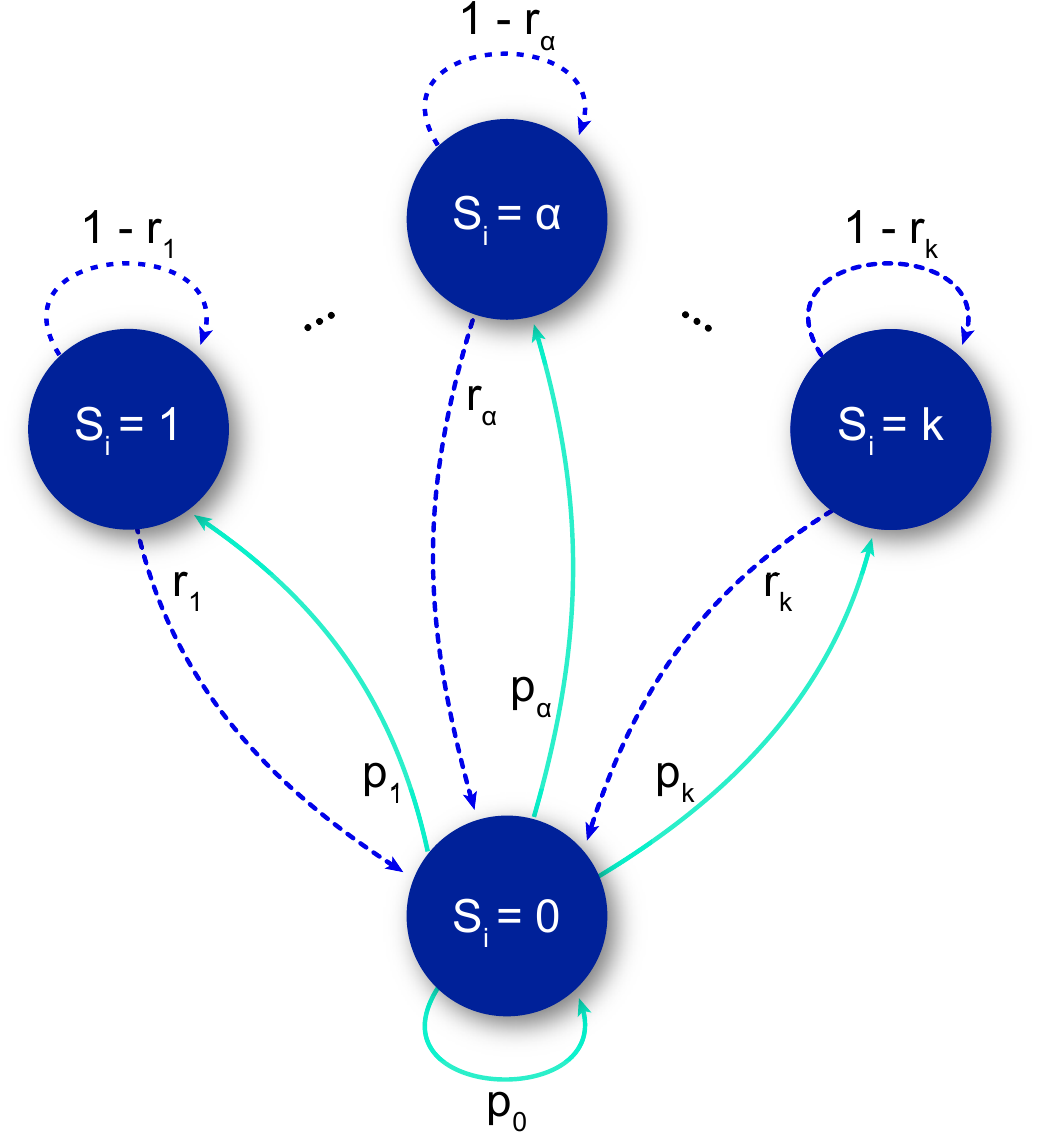}
    \caption{Schematic representation of the transitions occurring between the commitment states of the LES model.
Agents transition from being uncommitted ($s_i = 0$) to being committed to option $\alpha$ ($s_i = \alpha$) with a probability denoted as $p_{\alpha}$, where $\alpha$ ranges from 1 to $k$. If they fail to commit to any of the $k$ options, they remain uncommitted with a probability $p_0$. Once an agent is in a committed state, it can transition back to the uncommitted state with a probability $r_{\alpha}$. Notice that direct transitions between different committed states are not permitted.}
    \label{fig:model_rep}
\end{figure}

Building on recent advances in collective decision-making models inspired by honeybee behavior, we show how the LES model adheres to Weber's law of perception~\cite{reina_psychophysical_2018, pais2013}. In the context of decision-making, Weber's law states that the smallest detectable difference between two options is proportional to the magnitude of their average quality. This means that as stimulus intensity increases, a greater difference is needed to perceive a change. The law, which applies to various sensory modalities such as vision, hearing, and touch, illustrates the logarithmic nature of perception, where sensitivity diminishes as stimulus intensity rises. Remarkably, our findings reveal that varying levels of social feedback, exploration, and system size can influence the minimum noticeable quality difference in collective decision-making.  

Finally, we demonstrate that a limiting case of the LES model maps directly onto the well-known contact process. Our results near this limit reveal how the bio-inspired system reaches maximum consensus at the onset of critical behavior, illustrating how such systems may exploit criticality to enhance consensus formation - similar to how many biological systems are hypothesized to operate near a critical point~\cite{mora_bialek_2011, MAMunoz_review2018, Chialvo2020}.

We approach the study of the LES model by analyzing its mean-field equations and complementing these results with simulations on fully connected networks and regular lattices in finite dimensions. These simulations allow us to explore the impact of finite size effects and examine convergence times, providing a more comprehensive understanding of the model's behavior in both idealized and realistic settings.

The paper is organized as follows. In Section~\ref{sec:modelling}, we provide a brief review of the model formulation and its analytical solution, highlighting results in specific limiting scenarios. Section~\ref{sec:results} presents a thorough analysis of the model and its significant findings. In Sections~\ref{sec:sym_disc} and~\ref{sec:asym_disc}, we initiate our investigation by examining the possibility of consensus formation and its characteristic convergence time within a binary decision context. 
In Section~\ref{sec:webers_law}, we show the model's adherence to Weber's Psychophysical Law of perception. Following this, in Section~\ref{sec:finite_size}, we explore the significance of finite size effects, which leads us to identify, in Section~\ref{sec:phase_trans_cp}, a correlation between the equal-quality and equal-discovery binary model and the contact process, highlighting its characteristic non-equilibrium critical behavior. Finally, we extend the model to incorporate more than two options in a multi-site scenario in Section~\ref{sec:more_sites}, before summarizing and concluding our study in Section~\ref{sec:conclusions}.

\section{Modeling fully-connected scout-bee-like swarms}
\label{sec:modelling}

List, Elsholtz, and Seeley introduced their honeybee-inspired collective decision-making model in~\cite{list_independence_2008}. This model encapsulates the primary mechanisms through which a swarm of honeybees achieves a collective decision. Honeybees possess the ability to independently discover potential nest site options and evaluate their quality. Subsequently, they communicate their findings to the rest of the swarm. Concurrently, bees can interact with their peers and opt to adopt an opinion already present within the group.

The LES model comprises a swarm of $N$ bees, denoted by $i = 1, ..., N$, and $k$ potential nest sites, denoted by $\alpha = 1, ..., k$. Each site is associated with an intrinsic quality, $q_{\alpha} \geq 0$, and an \textit{a priori} self-discovery probability $\pi_{\alpha} \geq 0$, which are considered fixed model parameters and do not change over time. Throughout the text, and without loss of generality, we opt to label the sites in ascending order of quality, such that $q_1 \leq ... \leq q_k$.

The decision process unfolds in discrete time steps. At any given time step, a bee can either be uncommitted, denoted by the state variable $s_i(t) = 0$, or committed to a specific option $\alpha$, indicated by $s_i(t) = \alpha$, and signaling its preference to its peers. Bees have the flexibility to transition between the uncommitted state and any committed state, but cannot directly switch between different committed states. The system's evolution is governed by a set of transition probabilities (${ p_{\alpha, t+1}}, {r_{\alpha, t+1}}$), represented by arrows in Fig.~\ref{fig:model_rep}, determining commitment and uncommitment events based on the system state at time $t$ for the subsequent time step $t+1$:
\begin{eqnarray}\label{eq:state_evolution}
    s_i(t)&=&0 ~ \xrightarrow{p_{\alpha, t+1}} ~ s_i(t+1)=\alpha \nonumber \\ 
    s_i(t)&=&\alpha ~ \xrightarrow{r_{\alpha, t+1}} ~  s_i(t+1)=0,
\end{eqnarray}

The commitment probabilities $\{p_{\alpha, t}\}$ are determined by the following expression:
\begin{equation}
    \label{eq:comm_prob}
    p_{\alpha, t+1} = (1-\lambda) \pi_{\alpha} + \lambda f_{\alpha, t} \text{.}
\end{equation}
The first term of this equation represents the likelihood that a scout bee independently discovers site $\alpha$, while the second term quantifies the probability that the bee commits to state $\alpha$ by following the advice of its peers: $f_{\alpha,t}$ denotes the fraction of agents promoting site $\alpha$ at time $t$. These terms are weighted by the interdependence parameter, $\lambda$. This parameter, ranging between 0 and 1, determines the extent to which bees depend on each other to reach their commitment decision. Vanishing interdependence ($\lambda \rightarrow 0$) signifies that bees will commit to a site solely based on independent exploration. 
Conversely, high interdependence ($\lambda \rightarrow 1$) means that bees 
will mostly consider their peers advertisement 
in order to make a commitment decision.
Probabilities must be normalized, so Eq.~\eqref{eq:comm_prob} must satisfy $\sum_{\alpha = 0}^{k} p_{\alpha, t} = 1$. It's important to note that we include the uncommitted state $0$ in the normalization condition. Additionally, the normalization condition implies that $\sum_{\alpha=0}^{k} f_\alpha = 1$, and consequently $\sum_{\alpha=1}^{k} \pi_\alpha \leq 1$.

The uncommitment transition probabilities ${r_{\alpha, t}}$ are inherently linked to the qualities of the sites. In the original LES model, transitions from committed to uncommitted states occur deterministically after a predetermined commitment time has elapsed. An agent commits to an option and advertises it for a fixed duration. 
To render the model's equations
mathematically tractable, T. Galla~\cite{Galla_2010} replaced this deterministic process by a stochastic process defined by the following rates:
\begin{equation}
    \label{eq:uncomm_rate}
    r_{\alpha} = q_0 \left[ \frac{\mu}{K} + \frac{1-\mu}{q_{\alpha}} \right] \text{,} \quad \alpha = 1,...,k .
\end{equation}
The parameter $\mu$ represents the extent to which bees independently assess the quality of a site. When $\mu = 0$, the duration of a dance relies solely on the quality of the site, indicating that bees independently evaluate the site's quality. Conversely, when $\mu = 1$, the site's quality becomes irrelevant, and bees predominantly advertise an option for a generic period of time, typically related to a new parameter $K$. This parameter $K$ is uniform for all sites, representing, for example, the maximum quality among all the options. The parameter $q_0$ ensures that $0 < r_{\alpha} \leq 1$ and represents the characteristic time scale of the problem. Here, we set $q_0=1$, and primarily focus on the case where $\mu=0$. Consequently, the average duration of the advertisement for site $\alpha$ is $1/r_{\alpha}$, which is proportional to $q_{\alpha}$.
The stochastic representation of these transitions still preserves the principal characteristic of the deterministic duration in the original model: the higher the quality of the state, the longer agents will remain committed and advertise for it.

The stochastic problem can be analyzed using a master equation, from which one can derive a set of nonlinear differential equations describing the evolution of the average fraction values $\langle f_{\alpha,t}\rangle$, henceforth referred to as the average dancing frequencies, for each state $\alpha$. By following the mathematical details outlined in~\cite{Galla_2010}, and assuming a fully connected, mean-field-like system, one can arrive at the following equations:

\begin{equation}
	\langle\dot{f}_{\alpha,t}\rangle = \langle f_{0,t} \rangle [(1-\lambda)\pi_{\alpha} + \lambda \langle f_{\alpha,t}\rangle] - r_{\alpha} \langle f_{\alpha,t}\rangle
	\label{eq:model_time_evo}
\end{equation}

\noindent where 
$\langle f_{0,t} \rangle = 1 - \sum_{\alpha = 1}^{k} \langle f_{\alpha,t}\rangle$. 
Equation~\eqref{eq:model_time_evo} can be readily integrated numerically, for example, using the Euler scheme.
%
Furthermore, an expression for the stationary points, denoted as $f_{\alpha}^{*}$, can be obtained by solving a system of $k$ coupled equations, which is derived by setting $\langle\dot{f}_{j,t}\rangle = 0$. By rearranging the resulting equations, one can derive expressions for the stationary values of the population of each site in terms of $f_0^{*}$:
\begin{equation}
    f_{\alpha}^{*} = \left[ \frac{r_{\alpha}}{f_0^*} - \lambda \right]^{-1} (1-\lambda) \pi_{\alpha} \quad \alpha = 1,...,k \text{.}
    \label{eq:fj_analytical_2}
\end{equation}
As formulated, the model can be analyzed through agent-based simulations or by numerically integrating Eq.~\eqref{eq:model_time_evo}. However, equations can be derived to determine the model's stationary points, from which analytical expressions can be obtained under certain specific limits. In the following subsection, we provide an overview of these analytical solutions.

\subsection{Stationary deterministic solutions}\label{sec:analytical_sol}

\begin{figure*}
    \includegraphics[width=.95\textwidth]{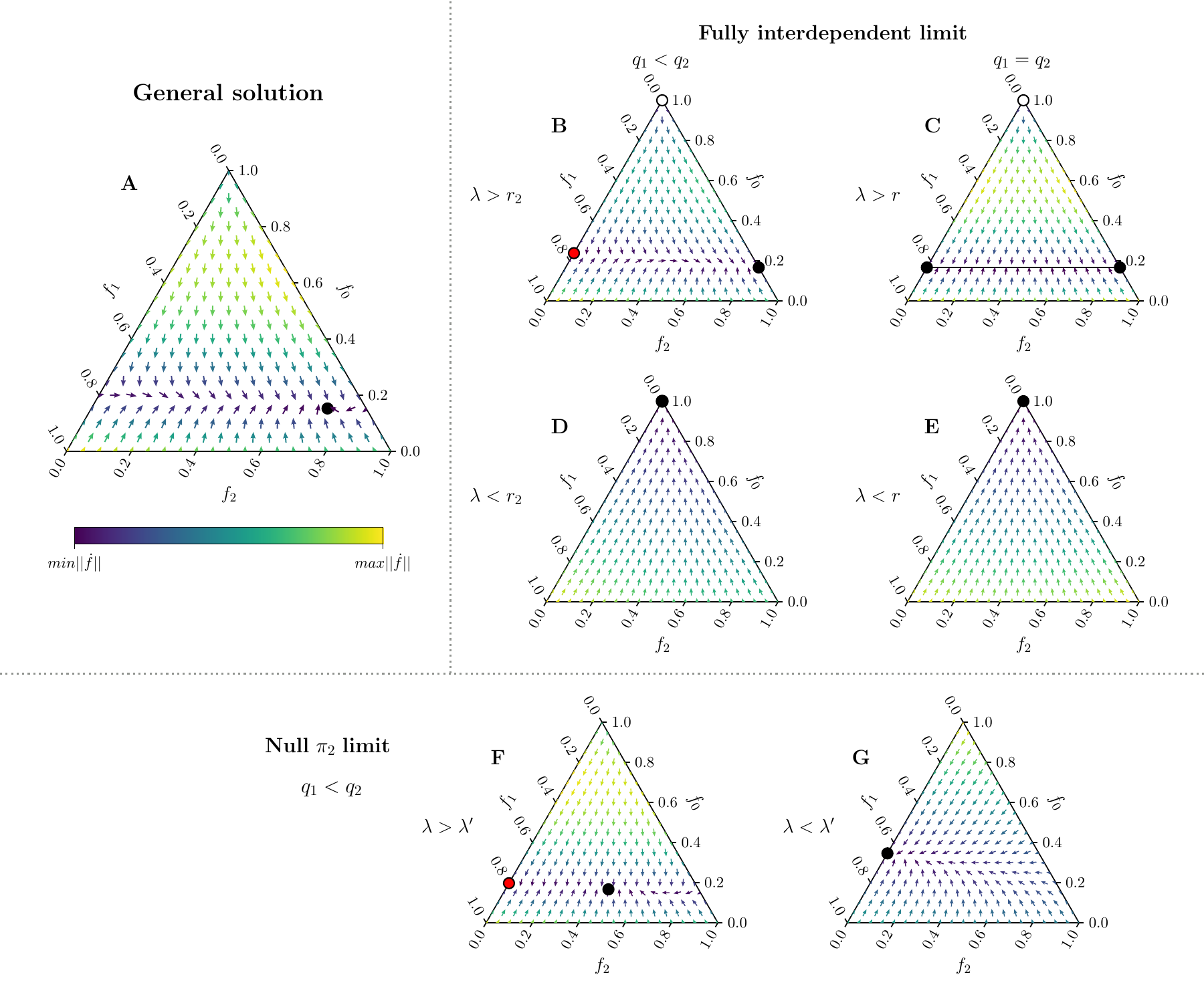}
    \caption{Flow diagrams depicting the dynamics of collective decision-making in a binary decision scenario. Arrows indicate the direction of the deterministic flow, while the color coding reflects the strength of the flow. Stable, saddle nodes, and unstable fixed points are denoted by black, red, and white circles, respectively. \textbf{A}: General solution for $\pi_1 = \pi_2 = 0.1$, $q_1 = 7$, $q_2 = 10$, $\lambda = 0.6$. Panels \textbf{B, C, D, E} illustrate solutions in the fully interdependent limit, $\pi_\alpha = 0 \; \forall \alpha$. In panels B and D, qualities are $q_1 = 7$, $q_2 = 10$, whereas in panels B and D, we have $q_1 = q_2 = 10$. Interdependence is $\lambda = 0.6$ in panels B and C, and $\lambda = 0.05$ in panels D and E. In both cases the absorbing transition appears at $\lambda = r_2, \; r_2 = 0.1$. \textbf{F, G}: Solutions in the limit of null discovery for the good option, $\pi_2 = 0$, while $\pi_1 = 0.25$. Qualities are $q_1 = 7$, $q_2 = 10$, and $\lambda = 0.6$ (Panel F) or $\lambda = 0.05$ (Panel G). The absorbing transition appears at $\lambda = \lambda' \simeq 0.43$. The analysis assumes perfect independent quality assessment, $\mu = 0$.}
    \label{fig:ternary_plots}
\end{figure*}

Since we can express $f_0^*$ as $f_0^*= 1 - \sum_{\alpha = 1}^{k} f_{\alpha}^*$, we can sum the $k$ equations given by Eqs.~\eqref{eq:fj_analytical_2} to derive a closed equation for the stationary value $f_{0}^*$:
\begin{equation}
    (1-f_0^*) \prod_{\alpha = 1}^k (r_{\alpha} - \lambda f_0^*) = (1-\lambda) f_0^* \sum_{\alpha = 1}^k \pi_{\alpha} \prod_{\mathclap{\substack{\beta = 1, \\ \beta \neq \alpha}}}^k (r_{\beta} - \lambda f_0^*)\text{.}
    \label{eq:f0_analytical}
\end{equation}

Eq.~\eqref{eq:f0_analytical} can be solved either numerically or rearranged into a $k+1$-th degree polynomial. In either approach, we find $k+1$ roots for $f_0$, some of which may lead to nonphysical solutions where $f_0^{*} > 1$. A linear stability analysis is provided in Appendix~\ref{sec:appendixA}.
In the specific case of $k=2$, assuming $q_2>q_1$ (or $r_2<r_1$), a single physical and stable solution $f^*_\alpha$ \; $\forall \alpha$ is obtained. An example of such a solution is depicted in Fig.~\ref{fig:ternary_plots}A, with values $r_2=1/10$, $r_1=1/7$, $\pi_1=\pi_2=0.1$, $\lambda=0.6$ and $\mu=0$.

\subsubsection{Limit of fully independent bees}

The limit case $\lambda = 0$ can be treated separately, as it permits a simpler analytical solution of Eq.~\eqref{eq:fj_analytical_2} (or, equivalently, Eq.~\eqref{eq:f0_analytical}), taking the form:

\begin{equation} \label{eq:f0_lambda0}
    f_0^{*} = \frac{1}{1 + \sum_{\beta=1}^k \frac{\pi_{\beta}}{r_{\beta}}}\text{.}
 \end{equation}
\begin{equation} \label{eq:fj_lambda0}
    f_{\alpha}^{*} = \frac{\pi_{\alpha}}{r_{\alpha}} \frac{1}{1 + \sum_{\beta=1}^k \frac{\pi_{\beta}}{r_{\beta}}} \mbox{\ \ } 1\leq \alpha\leq k.
\end{equation}

\subsubsection{Limit of fully interdependent bees}\label{sec:fully_interdep_lim}

In the limit cases where $\pi_{\alpha} = 0$ for all $\alpha$, or $\lambda = 1$, where individual sourcing of information has no influence, simpler analytical solutions exist. It should be noted that in dynamics originating from specific initial conditions, such as an all uncommitted population ($f_0(t=0) = 1$), committed populations never have the opportunity to grow in this limit. Beyond this particular case, we can derive the stable analytical fixed point of the model in this limit. The $k+1$ fixed points solutions of this limit take the form 
\begin{eqnarray}\label{eq:f0_pi_0}
    & f_0^* = 1, \quad & f_1^* = ... = f_k^* = 0 \\ 
    \text{and} & & \nonumber \\
    & f_0^* = r_{\beta}/\lambda, \quad & f_{\alpha}^* = (1 - r_{\beta}/\lambda)\delta_{\alpha,\beta}, \nonumber
\end{eqnarray}
where $\delta_{\alpha,\beta}$ is the Kronecker delta, and $1\leq \alpha,\beta\leq k$. In the case where $\lambda = 1$, the second family of solutions further simplifies to $f_0^* = r_{\beta}$, $f_{\alpha}^* = (1 - r_{\beta})\delta_{\alpha,\beta}$, $\forall \beta$. 

Although all these solutions are physically valid, only two of them (one in the case $\lambda = 1$) are stable.
Using linear stability analysis (LSA), we determine that the stability threshold between these two solutions is determined by the highest site quality, denoted as site $k$. When $\lambda < r_k$, the stable solution is the absorbing state $f_0^* = 1$. Thus, in the absorbing state, the system remains fully uncommitted. Conversely, when $\lambda > r_k$, the stable solution is the one with the best site taking the lead, with $f_{\alpha}^* = (1 - \frac{r_k}{\lambda})\delta_{\alpha,k}$, and the remaining proportion of uncommitted population being $f_0^* = \frac{r_k}{\lambda}$. In the special case where we set $\lambda = 1$, the stability threshold vanishes because $\lambda=1 \geq r_k$, and the only stable solution remaining is $f_0^* = r_k$, $f_{\alpha}^*= (1 - r_k)\delta_{\alpha,k}$. Details on the LSA analysis are provided in Sec.~\ref{sec:appendixB}.

In Fig.~\ref{fig:ternary_plots}B,D, we present a summary of the findings for the binary problem with $k=2$. When $\lambda \geq r_2$, only the fixed point where the best option is imposed remains stable, while the one where the inferior option is imposed acts as a saddle node (stable on the simplex $f_2 = 0$). As $\lambda$ decreases, these points gradually move upward (reducing either $f_1$ or $f_2$) until they merge with the point $f_0 = 1$ at $\lambda = r_{\alpha}$. In the regime where $\lambda < r_2$, the only remaining stable fixed point is the absorbing state $f_0^* = 1$ fixed point.

\subsubsection{Limit of equal-quality sites} \label{sec:equal_qualities_sol}

So far, our analysis has assumed that all sites differ in quality, or at least one site's quality is greater than the others.
However, when all qualities are equal, implying $r_{\alpha} = r$ for $1 \leq \alpha \leq k$, the only parameters that can break the symmetry between the populations of the different states are their discovery probabilities, $\pi_{\alpha}$.
If, on top of identical qualities, we also have identical discovery probabilities, $\pi_{\alpha}=\pi$ for $1 \leq \alpha \leq k$, the system will end up in a deadlock between the $k$ possible nest-site options. In other words, $f_1^* = \ldots = f_{k}^*$, and therefore, no consensus is reached.
In this particular limit though, the general solution for $f_0^*$ reduces to a second-degree polynomial, with the physical and stable solution being $f_0^* = (A - \sqrt{A^2 - 4\lambda r})/(2\lambda)$, where $A = k \pi (1-\lambda) + \lambda + r$.

In the specific case where we have equal-quality sites and fully interdependent bees ($\pi_\alpha = 0$ for all $\alpha$), we find two solutions $f_0^* = 1$ and $f_0^* = r/\lambda$ delimited by the stability threshold value $\lambda = r$. In the former solution, all the other frequencies vanish, i.e. $f_{\alpha}^* = 0$ for $1 \leq \alpha \leq k$, whereas in the latter the actual values of $f_{\alpha}^*$ remain undetermined (note that a null denominator appears in Eq.~\eqref{eq:fj_analytical_2}). According to LSA, any combination that satisfies $\sum_{\alpha=1}^k f_\alpha = 1 - r/\lambda$ is a potential solution. 
Consequently, the stationary state reached after numerically integrating Eq.~\eqref{eq:model_time_evo} is highly sensitive to the initial conditions. Any initial state satisfying $\sum_{\alpha=1}^k f_\alpha = 1 - r/\lambda$ will remain unchanged, while other initial conditions that do not satisfy this condition will evolve toward stationary points where one option dominates, and the others vanish. The specific option that prevails depends on which one initially has an advantage. When performing numerical simulations, finite-size effects further influence the decision outcome (see Secs.~\ref{sec:finite_size} and \ref{sec:phase_trans_cp}).

In Figs.~\ref{fig:ternary_plots}C and E, we summarize the results once more for the binary option case in the fully interdependent limit. When $\lambda > r$, all points in the simplex $f_1 + f_2 = 1 - r/\lambda$ (depicted as a black line) are plausible solutions. As $\lambda$ decreases, these solutions move towards lower values of $f_1$ or $f_2$ until they coalesce with the absorbing state $f_0^* = 1$ when $\lambda < r$.

\begin{figure*}[t!]
    \includegraphics[width=1.0\textwidth]{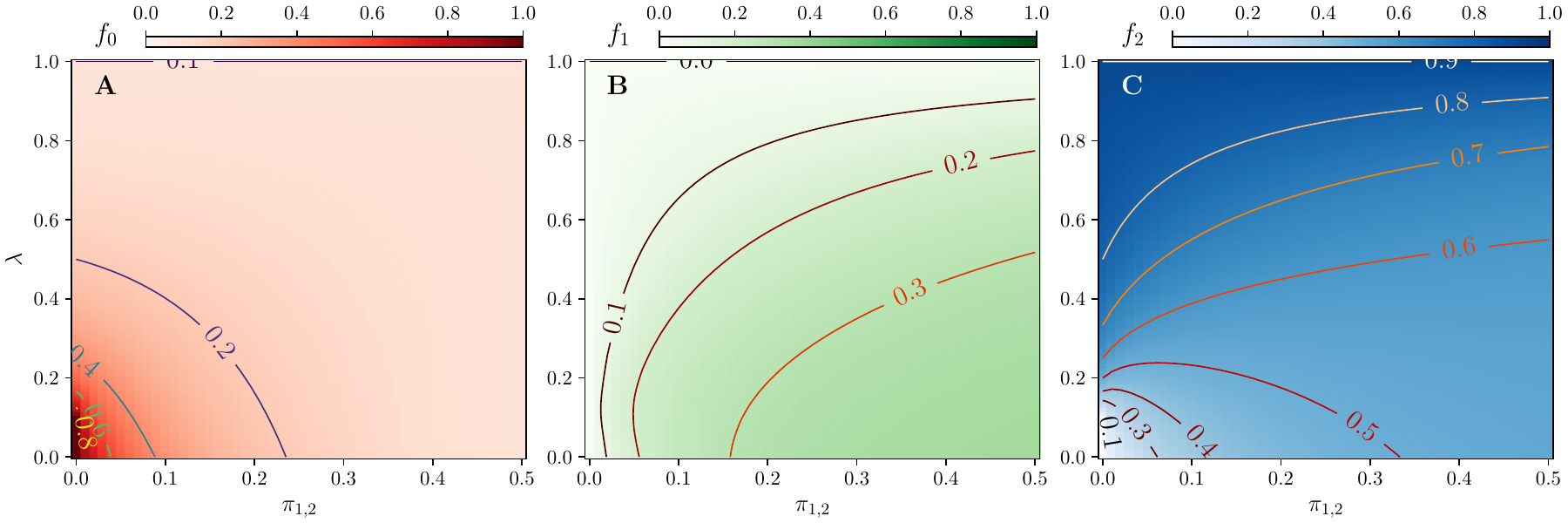}
    \caption{Color-maps illustrating the stationary values of $f_0$, $f_1$, and $f_2$ in the parameter space $(\pi_{1,2}, \lambda)$ are presented in panels \textbf{A}, \textbf{B}, and \textbf{C}, respectively. The site qualities are held constant at $(q_1, q_2) = (7, 10)$. Color gradients are used to indicate the quantitative values of $f_\alpha$ based on the color bars, while continuous lines represent iso-population levels in the parametric space.
    }
     \label{fig:sym_pi_state_space}
\end{figure*}

\subsubsection{Null discovery probability of the best site}

The null discovery probability of the best site, $\pi_k = 0$, defines a limit where the best option is never discovered independently. This limit emphasizes the significance of social influence in decision-making processes, particularly when the inherent quality of options is not apparent to individuals through independent assessment. This specific limit can also be examined analytically. Since $\pi_k = 0$, the differential equation for the best quality site simplifies to a straightforward expression, yielding two stationary solutions: $f_0^* = r_k/\lambda$ or $f_k^* = 0$. With the first solution, one can readily derive the remaining stationary frequencies:

\begin{eqnarray}\label{eq:pikeq0_fkneq0_sol}
     & & f_0^* = \frac{r_k}{\lambda}, \quad \nonumber f_{\alpha}^* = \frac{\pi_{\alpha} (1-\lambda) r_k}{\lambda (r_{\alpha} - r_k)} \qquad \alpha = 1,...,k-1, \\
     & & f_k^* = 1 - \frac{r_k}{\lambda}\left[ 1+ (1-\lambda)\sum_{\alpha = 1}^{k-1}\frac{\pi_{\alpha}}{r_{\alpha} - r_k}\right]\text{.} \nonumber
\end{eqnarray}

For the second stationary point $f_k^* = 0$, one arrives at a system of $k-1$ coupled equations for the other committed populations. 
While it is possible to derive analytical expressions for $f_{\alpha}^*$ ($\alpha \neq k$) and $f_0^*$ for a small number of sites, it is often more practical to use Eqs.~\eqref{eq:fj_analytical_2} and ~\eqref{eq:f0_analytical}, or to resort to numerical integration of Eq.~\eqref{eq:model_time_evo} to obtain the complete solution.

In conclusion, the noteworthy aspect of this particular limit is the transition from a stable state where there is no population for the best option ($f_k^*=0$) to a state where $f_k^* \neq 0$, as described by Eq.~\eqref{eq:pikeq0_fkneq0_sol}. The transition occurs as $\lambda$ increases, at the specific value 
\begin{equation}
    \lambda' = \frac{1+\psi}{r_k^{-1}+\psi} \qquad \text{where} \; \psi = \sum_{\alpha = 1}^{k-1}{\frac{\pi_{\alpha}}{r_{\alpha}-r_k}}
    \label{eq:null_pi2_lambda_crit}
\end{equation}
It should be noted that while conducting a linear stability analysis on these solutions confirms the change in stability of the solution with $f_k^* = 0$ (stable when $\lambda < \lambda'$), the solution provided by Eq.~\eqref{eq:pikeq0_fkneq0_sol} is always stable. However, it becomes nonphysical when $\lambda < \lambda'$, as $f_k^* < 0$. The change in sign occurs precisely at the same value $\lambda'$. When the qualities are equal, the regime where $f_k^* \neq 0$ disappears, effectively reducing the system to one with only $k-1$ options, which can be analyzed using the general solution.

The two regimes are depicted in Figs.~\ref{fig:ternary_plots}F and G, focusing on the specific case of $k=2$. As $\lambda$ approaches $\lambda'$ from above, $f_2 \rightarrow 0$ and the stable point approaches the unstable solution (Fig.~\ref{fig:ternary_plots}F). At $\lambda=\lambda'$, these two points merge, and in the regime $\lambda < \lambda'$, the only stable solution is where no population is committed to the best-quality site.

\section{Results}
\label{sec:results}

Going beyond the mean-field approach, previous studies have explored the 
stochastic dynamics of this decision-making model through simulations  
on a regular lattice~\cite{Galla_2010} and on random networks~\cite{marchpons2024_kilobots}. 
Additionally, experiments have been conducted using mini-robots as a physical platform~\cite{marchpons2024_kilobots}, albeit with a limited parameter set explored. 
In these random scenarios, the deviation from mean-field behavior is primarily influenced by network connectivity. As analyzed in~\cite{marchpons2024_kilobots}, 
the emergence of a giant component within the communication network is crucial for the system to transition toward mean-field-like results.


In the following, we focus on characterizing the model's behavior using a 
mean-field approach, conducting an exhaustive exploration across a broad 
parameter range. 
To quantify the stationary results, we rely on analytical expressions. 
Simulations conducted on fully connected systems or square lattices 
yield average results that align perfectly with these analytical solutions. 
However, to complement our analysis, we take advantage of simulations to measure 
the time needed to reach the stationary state and to address finite-size effects.

We will first concentrate on the specific case of a binary decision problem, 
a topic extensively studied in various general opinion dynamics models~\cite{galam2002, galam2008, castellano_statistical_2009, StarniniJSTAT2012, valentini2017, redner_reality-inspired_2019, VazquezPRE2019, RamirezPRE2022, reina_voter_2023, RamirezPRE2024}. 
This scenario has also been the focus of other honeybee-inspired models~\cite{valentini2014, reina_desing_pattern, reina_model_2017, gray_multiagent_2018}. 
The generalization to larger values of $k$ is discussed later in Section~\ref{sec:more_sites}.

\begin{figure*}
    \centering
    \includegraphics[width=1.0\textwidth]{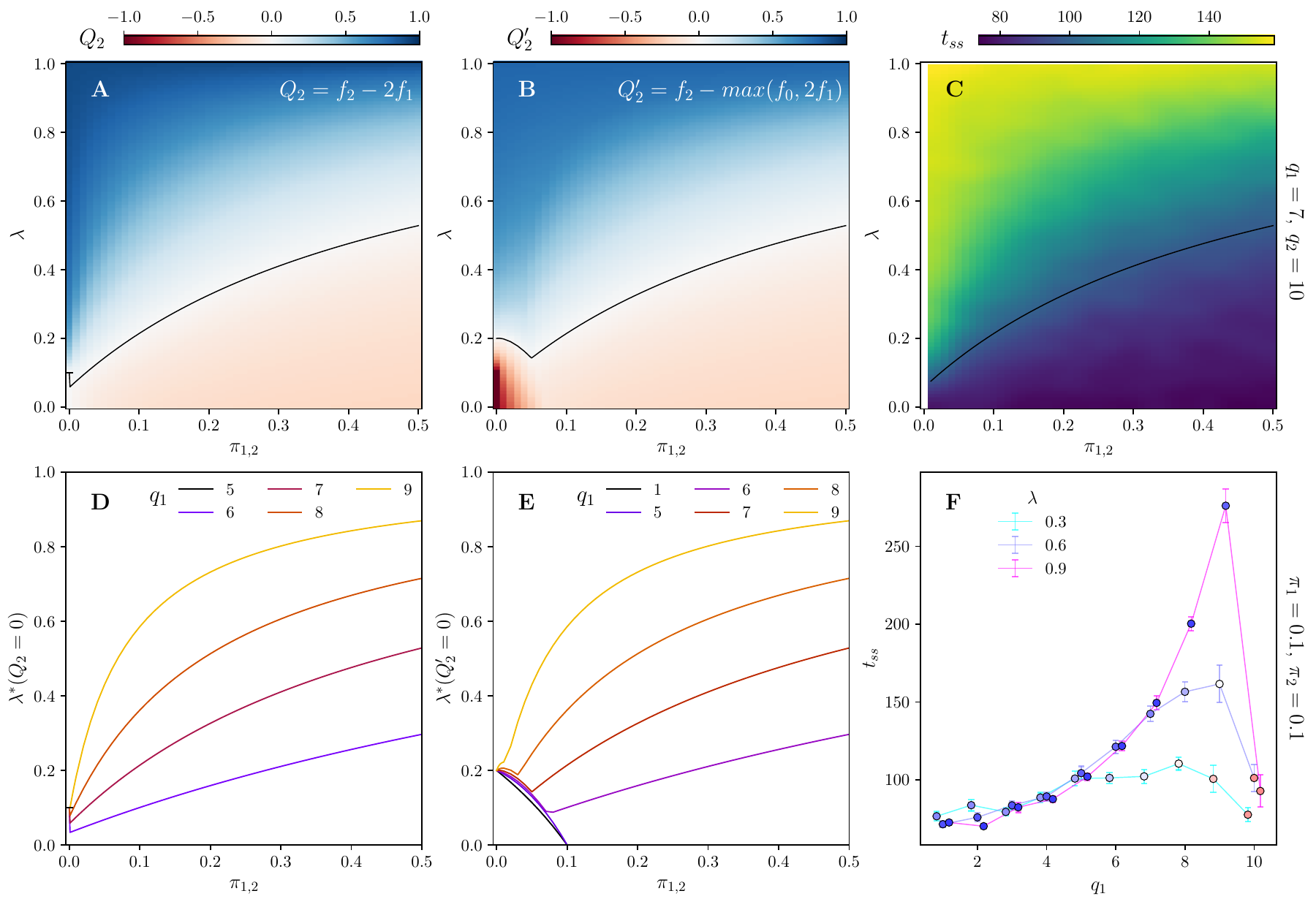}
    \caption{\textbf{A},\textbf{B}: Color-maps illustrating the consensus parameters $Q_2$ and $Q_2'$, as defined in Eqs.~\eqref{eq:Qdefinition} and \eqref{eq:Qalt_definition} respectively with $x=2$, across the parameter space $(\pi_{1,2},\lambda$). The black lines represent the crossover points $Q_2=0$ or $Q_2'=0$. \textbf{C}: Time to reach the stationary state, $t_{ss}$, obtained from simulations with a swarm size of $N=5000$ bees. The fully interdependent limit ($\pi_{\alpha} = 0$ or $\lambda = 1$) is not included in the maps (see discussion in Sec.\ref{sec:phase_trans_cp}). The site quality parameters are held constant at $(q_1, q_2) = (7, 10)$ across all three maps. \textbf{D},\textbf{E}: Crossover lines $\lambda^* (Q_2, Q_2'=0)$ as a function of $\pi_{1,2}$. Each curve represents a different value of $q_1$, while $q_2=10$ is held constant. \textbf{F}: Time to reach the stationary state, $t_{ss}$, obtained from simulations plotted against the quality of the inferior option, $q_1$, while maintaining  $q_2=10$, for three values of the interdependence. 
    Parameters $\pi_1 = \pi_2 = 0.1$ and $N=5000$ are fixed. Simulations start from uncommitted initial conditions.
    The markers are color-coded based on the consensus established at the stationary state. They are slightly jittered around the $q_1$ axis to allow for better visualization of the color code.}
    \label{fig:sym_Q_and_time}
\end{figure*}

As detailed in the model description section, we consistently assume that site 2 
holds the highest quality, denoted by $q_2 > q_1$. 
Additionally, we assume that bees independently assess the qualities of the sites, 
represented by the model parameter $\mu = 0$. 
This implies that the abandonment rates of the dances depend solely on the 
qualities of each site, defined as $r_{\alpha} = q_0 / q_{\alpha}$. 
Previous research on this model has shown that, under these conditions, 
the swarm can identify the best site across a wide range of 
interdependence values, $\lambda$ \cite{list_independence_2008, Galla_2010}.
In this setup, we investigate the interplay between group communication, 
represented by the model parameter $\lambda$, and individual exploration success, 
captured by $\pi_\alpha$. 
We explore scenarios where sites may vary in their likelihood of being discovered, 
denoted by $\pi_1 \neq \pi_2$ (asymmetric scenario), or where they share the same 
discovery probability, denoted by $\pi_1 = \pi_2$ (symmetric scenario).

\subsection{Symmetric discovery scenario}\label{sec:sym_disc}

We start by examining the case where the available sites have an equal likelihood of being found, 
i.e., $\pi_1 = \pi_2 = \pi_{1,2}$, and they only differ in their quality, with $q_1 < q_2$.
We present our findings by exploring the state space defined by $(\pi_{1,2}, \lambda)$ for 
various combinations of site qualities, while keeping $q_2 = 10$ constant and varying $q_1$.

Figure~\ref{fig:sym_pi_state_space}A-C represents the particular values of the 
frequencies $f_0$, $f_1$ and $f_2$ for a particular choice of the qualities, 
$(q_1, q_2) = (7,10)$. 
In these plots, we observe smooth trends for all these quantities. 
As the interdependence increases, the frequencies $f_0$ and $f_1$ decrease 
regardless of the value of the discovery probabilities. 
This decline favors the proportion of bees committed to the best site, $f_2$. 
When bees rely more on the opinions of their peers, the best quality option 
benefits from the positive feedback generated by longer advertisement times. 
Conversely, when independent discoveries increase, $f_1$ increases at the 
expense of $f_0$ and $f_2$. 
This shift occurs because as $\pi_{1,2}$ values increase while keeping $\lambda$ 
constant, more advertisements are motivated by independent discoveries rather 
than by opinion sharing. Consequently, information is introduced equivalently 
for both sites, and the best-quality reinforcement effect of $\lambda$ is hindered. 
In the limit where the discovery probabilities are small, the good option is easily 
imposed by interdependence-mediated discussion. However, when $\pi_{1,2} = 0$, 
we observe a transition to the absorbing regime, $f_0^*=1$, at $\lambda = r_2=0.1$.

With these results in mind, we can quantify the outcome of the decision process by defining a consensus. This involves setting a threshold condition on the values of the frequencies $f_{\alpha}$ that is necessary to conclude that the swarm has reached a decision. Such a consensus will be defined with respect to the best option, as our main focus is discerning whether the swarm is able to retrieve the best available option.

The simplest approach is to define a threshold on the value of the population committed to the best option. For instance, one can require that at least half of the total population is committed to the best option in order to consider the system to have made a successful decision. This threshold line is depicted as the contour line $f_2 = 0.5$ in Fig.~\ref{fig:sym_pi_state_space}C. We observe that a large portion of the state space $(\pi_{1,2}, \lambda)$ lies above this threshold line. When the interdependence is low, and thus the competition between options is stronger, the system may fail to achieve such a consensus. Furthermore, when the discovery probabilities decrease drastically, the system remains mostly uncommitted, which is insufficient to reach a consensus.
Increasing the threshold value for $f_2$, such as requiring a two-thirds majority ($f_2 = 2/3$), will shift the crossover point towards higher values of interdependence.

While setting a threshold value on one of the populations provides a straightforward measure of consensus, this  method may fail to capture scenarios where the population committed to the best option is substantial, yet faces significant competition from other sites. Additionally, it might overlook situations where one site clearly leads, but the committed population has not yet met the prescribed threshold.
To address these concerns, we adopt the consensus definition proposed by LES in their original paper ~\cite{list_independence_2008}. This approach involves comparing the two largest values of $f_{\alpha}$, offering a more comprehensive assessment of the decision-making dynamics. In a general manner, we can define a consensus measure as follows:
\begin{equation}\label{eq:Qdefinition}
    Q_x = f_2 - x f_1 \text{.}
\end{equation}
where $x$ indicates whether we require a simple majority ($x=1$), a two-thirds majority ($x=2$) provided that $f_0=0$, or any other desired threshold between the two best options. Figure \ref{fig:sym_Q_and_time}A shows the value of $Q_2$ in the $(\pi_{1,2}, \lambda)$ space for the same choice of qualities as before. The state space is divided between a region of consensus for high $\lambda$ and region without consensus for small $\lambda$, separated by a consensus crossover line, $Q_2(\lambda^*)=0$.
We observe a similar trend as with the $f_2$-threshold consensus definition: high values of $\pi_{1,2}$ require high values of the interdependence in order to retrieve the best option among the \textit{noise}, which is equivalently introduced for both options. On decreasing the discovery probabilities, the range of $\lambda$ yielding consensus grows, due to the main mechanism under the collective decision being the positive reinforcement on the best quality options through interdependence.

Fig.~\ref{fig:sym_Q_and_time}D illustrates the consensus crossover lines for different choices of $q_1$, while keeping $q_2=10$ fixed. When we increase the quality of the bad option, consequently increasing its advertisement time, the system requires a higher degree of interdependence to favor the higher quality option. This trend would reach a turning point when $q_1 = q_2$. In this scenario, the system reaches a deadlock, where $f_1 = f_2$. Conversely, when $q_1$ is reduced, the crossover line shifts to lower values of $\lambda^*$, potentially leading to a situation where interdependence is not necessary to achieve consensus. For instance, the crossover line for $q_1 = 5$ lies along the x-axis, with $\lambda^*= 0$ for all $\pi_{1,2}$. In this case, the difference in advertisement time is sufficient to maintain a significant difference between $f_2$ and $f_1$, facilitating consensus even in the absence of interdependence.
Furthermore, it's notable that the consensus crossover lines terminate abruptly at the point $(\pi_{1,2}, \lambda) = (0.0, 0.1)$. As we demonstrated earlier, in the fully interdependent limit ($\pi_{1,2} = 0$) the system stays fully uncommitted until the threshold $\lambda = r_2=0.1$. Beyond this threshold, $f_2$ begins to take positive values while $f_1$ remains zero, resulting in trivial consensus.

\begin{figure*}[t!]
    \centering
    \includegraphics[width=\textwidth]{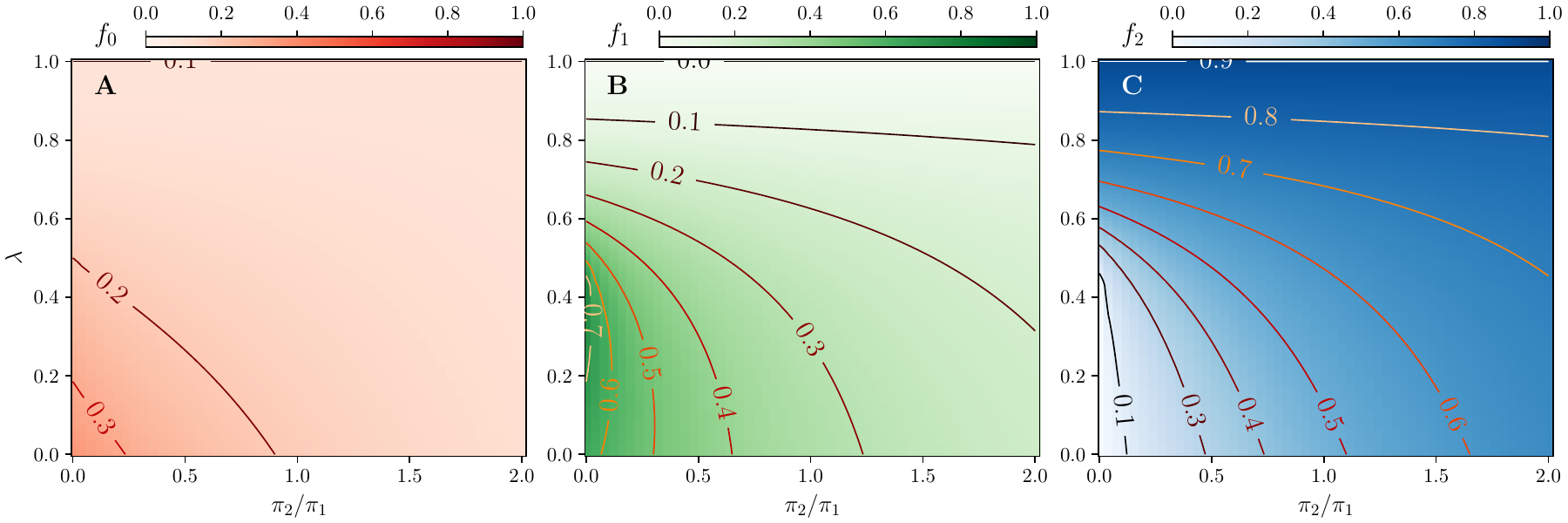}
    \caption{Color-maps illustrating the stationary values of $f_0$, $f_1$ and $f_2$ are presented in panels \textbf{A}, \textbf{B}, and \textbf{C}, respectively, for the asymmetric discovery scenario in the parameter space $(\pi_2/\pi_1, \lambda)$.
    The site site qualities are fixed at $(q_1, q_2) = (7, 10)$, and the discovery probability for site 1 is set to $\pi_{1} = 0.25$, while $\pi_2$ is changed. 
    Color gradients indicate the quantitative values of $f_\alpha$ according to the color bars, while continuous lines represent iso-population levels in the parametric space.}
    \label{fig:asym_pi_state_space}
\end{figure*}

For $\pi_{1,2} > 0$, we can predict the value of $q_1$ below which consensus will always be achieved without the need for interdependence. Using the model solution at $\lambda = 0$ and the general definition of consensus, $Q_{x} = f_2 - x f_1$, we arrive at the following condition:
\begin{equation*}
    q_1 < \frac{q_2}{x}
\end{equation*}
Consequently, if $x = 2$, consensus is readily achieved if $q_1 < q_2 / 2$, while a simple majority consensus is always guaranteed if $q_1 < q_2$. Reintroducing the independent quality assessment parameter of the model, $\mu$, modifies the values of these thresholds. We provide a brief discussion of this case in Appendix \ref{app:mu}.

We would also like to acknowledge that the definition of consensus in Eq.~(\ref{eq:Qdefinition}) may not be very appropriate to describe the low discovery region of the parameter space where $\pi_{1,2} \rightarrow 0$. There, we observe that $\lambda^*$ decreases, suggesting that a scarce environment with low $\pi_{1,2}$ is the most beneficial situation for the swarm, as it allows for the widest range of $\lambda$ yielding consensus. However, it's important to note that the range of uncommitted population increases drastically in that region, and the system can even transition to an absorbing state around $\lambda = r_2$. Therefore, this effect should be taken into account in the quorum measure. In the original paper by LES~\cite{list_independence_2008}, the consensus definition is strengthened by requiring a minimum proportion of agents engaged in the decision process. Here, we propose a slightly different definition to encompass this effect along with the site competition.
\begin{equation}\label{eq:Qalt_definition}
    Q_x' = f_2 - \max(f_0, xf_1)\text{.}
\end{equation}

\noindent Consequently, when there is not a sufficient population committed to an option other than the winner, the competition is solely against the undecided population. Unlike when two options compete, where remaining unresolved is not considered a valid outcome of the decision process, this definition only requires that the committed population exceeds the uncommitted to establish a quorum.
As shown in Fig.~\ref{fig:sym_Q_and_time}B (with $x=2$), this definition automatically excludes the region of small $\pi_{1,2}$ and $\lambda$ from consensus. This ensures that we do not conclude there is consensus when a large proportion of the population remains uncommitted, while still capturing the effect of site competition that occurs when the discovery probabilities increase.
Note that under this definition, the condition that consensus is granted if $q_1 < q_2/x$ no longer holds (Fig.~\ref{fig:sym_Q_and_time}E), at least not for all $\pi_{1,2}$. When the subleading population is uncommitted, we find the threshold value $\pi_2 > r_2$. Consequently, we observe a modified crossover line from $\pi_{1,2} = 0$ up to $\pi_{1,2} = r_2$. Right at the fully interdependent limit ($\pi_{1,2} =0$), the condition that must be satisfied is $\lambda > 2 r_2 = 0.2$.

\begin{figure*}[t!]
    \centering
    \includegraphics[width=\textwidth]{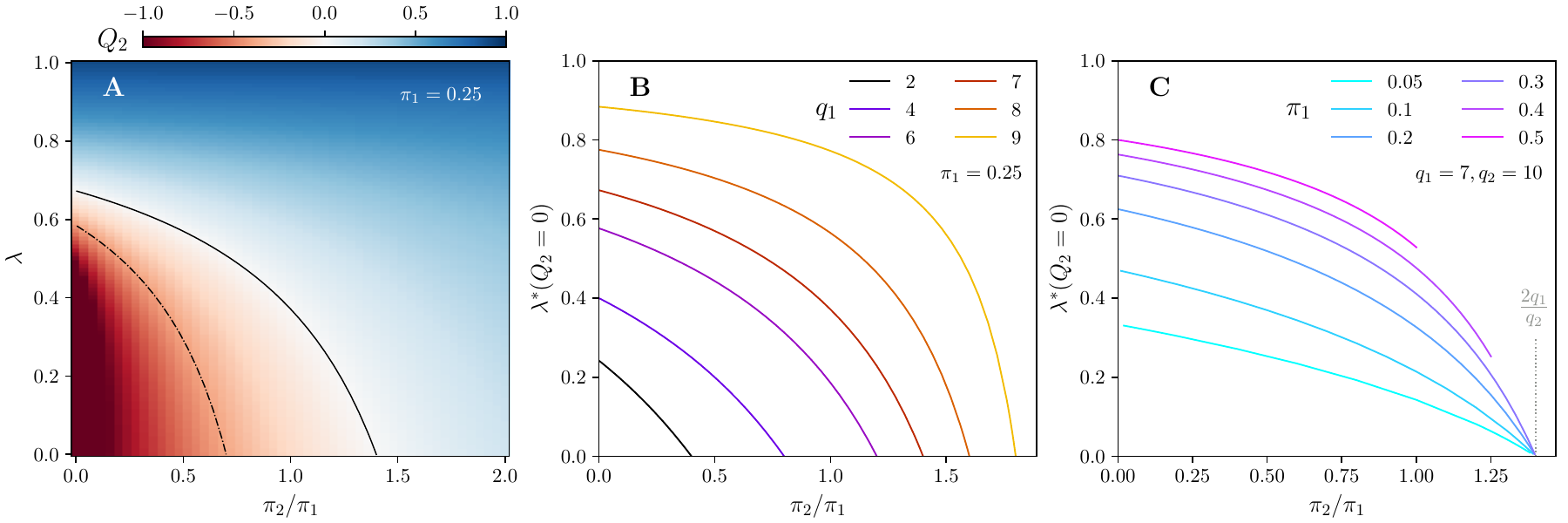}
    \includegraphics[width=.66\textwidth]{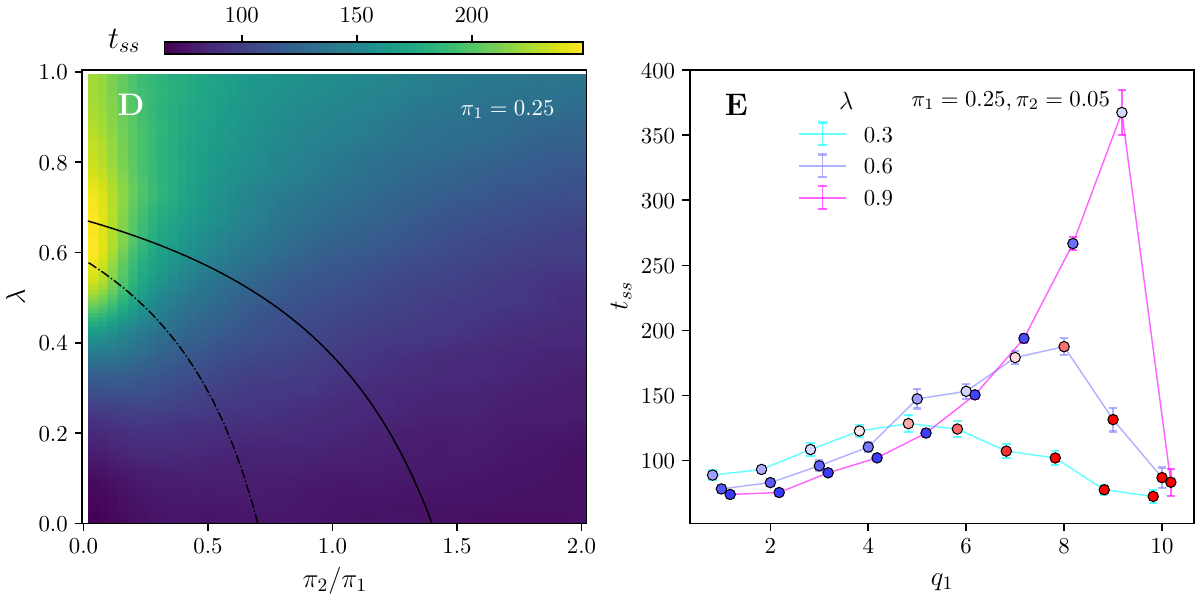}
    \caption{\textbf{A}: Color-map illustrating the stationary value of the consensus parameter $Q_2$ (as defined from Eq.~\eqref{eq:Qdefinition} with $x=2$) in the parametric space $(\pi_2/\pi_1, \lambda)$. 
    Site qualities are fixed at $(q_1, q_2) = (7, 10)$, and the discovery probability for site 1 is set to $\pi_1 = 0.25$. The solid line indicates $Q_2=0$, while the dashed line indicates a simple majority, $Q_1 = 0$.
    \textbf{B}: Crossover lines $\lambda^*$ above which the consensus $Q_2$ becomes positive as the self-discovery probability $\pi_{2}$ is varied. Each curve represents a different value of site-1 quality $q_1$, while $q_2=10$ and $\pi_1 = 0.25$ are held constant.
    \textbf{C}: Crossover lines $\lambda^*$ above which $Q_2$ becomes positive as $\pi_2$ is varied. In this panel, qualities are held constant, $(q_1, q_2) = (7, 10)$, and we illustrate the effect of different values of $\pi_1$. \textbf{D}: Time to reach the stationary state $t_{ss}$ in the state space ($\pi_2/\pi_1, \lambda)$ obtained from simulations. Parameters $\pi_1 = 0.25$, $(q_1, q_2) = (7, 10)$ and $N = 5000$ are fixed. \textbf{E}: Time to reach the stationary state, $t_{ss}$, obtained from simulations plotted against the quality of the inferior option, $q_1$, while maintaining $q_2 = 10$, for three values of the interdependence. Parameters $(\pi_1, \pi_2) = (0.25, 0.05)$ and $N = 5000$ are fixed. Simulations start from uncommitted initial conditions.
    Color markers indicate the consensus value attained at the stationary state. They are slightly jittered around the $q_1$ axis to allow for better visualization of the color code.
    }
    \label{fig:asym_pi_Q_lines_and_time}
\end{figure*}

Up until now, we have focused on the swarm's ability to reach consensus across different combinations of model parameters. However, it's equally important to consider the decision time, whether consensus is achieved or not, to fully understand the effect of these parameters on the decision process. Figure \ref{fig:sym_Q_and_time}C shows the timesteps required for the system to settle at the stationary state, obtained from simulations, across the state space $(\pi_{1,2}, \lambda)$ (details on how $t_{ss}$ is obtained are provided in the Appendix \ref{sec:appendixD}). 

The trend observed is similar to that of consensus: interdependence tends to increase the decision time across all values of $\pi_{1,2}$, while the discovery probabilities speed up the process, reducing $t_{ss}$. This correlation between the established consensus value and the time to reach the stationary state suggests a speed-accuracy trade-off, commonly observed in biological systems~\cite{chittka2009_speedacc} an in models of collective decision making~\cite{valentini2016, reina_voter_2023}. Achieving the best decision typically takes longer at higher values of $\lambda$, while processes are quicker below the consensus crossover line, where a less accurate consensus, such as a simple majority, can be reached much faster. 
Similarly, increasing the value of the discovery probabilities, and thus the introduction of independent information, accelerates the decision process at the cost of a lower value of $Q'$, which can be even negative, depending on $\lambda$. Only at the limits $\lambda \rightarrow 0$ or $\lambda \rightarrow 1$ $\pi_{1,2}$ has practically no effect: in the former, there's no discussion mediated by interdependence, resulting in a quickly established yet inaccurate stationary state, while in the latter, social feedback dominates the process, with little impact from individual exploration.
It's interesting to note the agreement between the stationary times discussed here and the relaxation times obtained from the linear stability analysis eigenvalues, as depicted in Fig.~\ref{fig:char_times}.

Finally, we must also acknowledge the influence of the site qualities on the stationary time. Figure \ref{fig:sym_Q_and_time}F illustrates $t_{ss}$ versus $q_{1}$, with fixed $q_{2}$ and discovery probabilities, for three values of the interdependence parameter. Each point is color-coded according to the actual value of the achieved consensus at the stationary state.
Primarily, we observe that as $q_1$ increases, the stationary time tends to increase, especially for large $\lambda$. This trend persists until the point where the system fails to reach consensus, at which point $t_{ss}$ starts to decrease. While this doesn't necessarily indicate a direct speed-accuracy trade-off (as $Q'$ was already decreasing before $t_{ss}$ started to do so), it's intriguing to note the correlation between $t_{ss}$ and the consensus value.
The speed-accuracy trade-off reemerges when examining fixed values of $q_1$: increasing $\lambda$ leads to higher consensus values, but also to longer stationary times. This trend holds for moderate values of $q_1$; when the problem difficulty is low ($q_1 << q_2$), $\lambda$ has minimal impact on $t_{ss}$.

\subsection{Asymmetric discovery scenario}\label{sec:asym_disc}

In the following section, we decouple the parameters $\pi_1$ and $\pi_2$, allowing one site or the other to be discovered more easily by the swarm. We still maintain $q_1 < q_2$. Consequently, we anticipate that when $\pi_2 > \pi_1$, the group will have no trouble reaching consensus, whereas when $\pi_2 < \pi_1$, interdependence will become crucial for the group to select the best possible option.

Continuing in a similar manner as the previous section, we explore the state space for a particular choice of site qualities. With three parameters -- $\pi_1$, $\pi_2$ and $\lambda$ --, we maintain $\pi_1$ fixed and present our results in the state space ($\pi_2/\pi_1, \lambda$). Figure \ref{fig:asym_pi_state_space} illustrates the behavior of $f_0$, $f_1$ and $f_2$ for site qualities $(q_1, q_2) = (7,10)$ and $\pi_1 = 0.25$. Once more, we observe a decrease in the proportion of the uncommitted population as either the discovery probability (for site 2) or the interdependence increases. This decrease predominantly benefits the population committed to option 2, as evidenced by the behavior of $f_1$ and $f_2$. As expected, increasing $\pi_2$ leads to a growth in $f_2$ while diminishing the other populations in the system. Similarly, raising the values of the interdependence also results in an increase in the population committed to the better option, accompanied by a decrease in $f_1$. It is noteworthy that even if the discovery probability of the better option is much smaller than that of the other option, sufficiently large values of interdependence enable the swarm to still select the better option. In this regime, only when interdependence is low does the worse option take the lead. In fact, for interdependence below a certain value $\lambda'$, the population committed to the better option is exactly null (as indicated by the dash-marker in Fig.~\ref{fig:asym_pi_state_space}). Similarly to the symmetric scenario, where we observed an absorbing transition at $\lambda = r_2$ in the fully interdependent limit ($\pi_{\alpha} = 0$), here, when $\pi_2 = 0$, we observe a transition from $f_2 = 0$ to $f_2 \neq 0$ at $\lambda = \lambda'$. This value of $\lambda'$ depends on both the qualities and the discovery probability of option 1.

As before, we can quantify the outcome of the decision process in terms of a consensus parameter. Requiring $f_2$ to reach a certain threshold value produces crossover lines $\lambda^*$ that decrease with $\pi_2$, as seen in the contour lines on Fig.~\ref{fig:asym_pi_state_space}C. This trend holds up to point where consensus is achieved without the need for interdependence, depending on the required threshold value. Beyond this point, the difference in advertising times is enough to impose a majority on the better option, regardless of the particular values of $\pi_1$ and $\pi_2$.

Introducing the consensus definition that accounts for the difference in the two most populated sites produces a very similar trend in the crossover lines. In Figure \ref{fig:asym_pi_Q_lines_and_time}A, the value of a two-thirds majority consensus in the voting population ($Q_2$) across the state space is presented, along with its crossover line (the solid line), as well as the crossover line for a single majority consensus, $Q_1$ (the dashed line). 
High interdependence facilitates consensus even in cases of drastic differences in the discovery probabilities, whereas at smaller values of $\lambda$, a region where consensus cannot be reached emerges. Below the $Q_1$ crossover line, the site with the lower quality is imposed, at least, by a simple majority.

Figure \ref{fig:asym_pi_Q_lines_and_time}B illustrates the crossover line $\lambda^*(\pi_2; \pi_1, x=2)$ for different choices of the bad site quality, $q_1$. The region where consensus cannot be reached expands as $q_1$ increases, pushing the parameters that yield consensus to very high levels of interdependence or a much greater probability to discover the good site. We can determine the value of $\pi_2$ from which consensus can be achieved without the need for interdependence. Using the solution at $\lambda = 0$ (Eqs. \eqref{eq:f0_lambda0} and \eqref{eq:fj_lambda0}), and the consensus definition, this value is given by:
\begin{equation*}
    \pi_2 \geq x \frac{\pi_1 q_1}{q_2}\textit{.}
\end{equation*}
As with the symmetric discovery scenario, introducing non-null values of $\mu$ will alter the value of this threshold; however, the behavior of the system remains unchanged. Further details are provided in Appendix \ref{app:mu}.

The overall magnitude of the discovery probabilities also influences the consensus crossover. Similar to the observations in the symmetric discovery scenario, smaller magnitudes of the individual $\pi$s necessitate smaller values of $\lambda^*$. This pattern persists in the asymmetric scenario as well. Figure \ref{fig:asym_pi_Q_lines_and_time}C illustrates this effect: even with the same ratio $\pi_1/\pi_2$, if the actual values of $\pi_1$ and $\pi_2$ are smaller, the consensus crossover line occurs at lower values of interdependence.

It's worth noting that unlike the symmetric discovery probability case studied previously, in this scenario and for the specified parameters, there is not a significant region where the uncommitted population grows noticeably, provided that one of the options has a sufficiently large discovery probability. Hence, there's no need to introduce a modified consensus definition that emphasizes the presence of a large uncommitted group. This feature only becomes relevant if, in the limit $\pi_2 \rightarrow 0$, either $\pi_1 \rightarrow 0$ --which results in the previously discussed scenario--, or $q_1 \rightarrow 0$, implying that option one is effectively expendable.

Finally, we turn our attention to the time required to reach the stationary state, as depicted in Fig.~\ref{fig:asym_pi_Q_lines_and_time}D. Once again, we observe the time dilation effect of $\lambda$, coupled with an increase in consensus, indicating a speed-accuracy trade-off. However, in contrast, when $\pi_1$ is held constant, increasing $\pi_2$ does not exhibit the same effect. Increasing the discovery probability for the good option simplifies the decision problem, achieving better consensus in a shorter time. Interestingly, we notice that in the limit $\pi_2 \rightarrow 0$, the stationary time increases abruptly around the consensus crossover. This implies that achieving consensus for the best option in such an unfavorable limit comes at the expense of a significantly prolonged decision process. 
A similar effect is observed when inspecting the characteristic relaxation times obtained from the solutions' LSA (see Fig.~\ref{fig:char_times}). 
There we observe that this increase in the relaxation time is instead found around the $f_2$ transition at $\lambda = \lambda'$.

When examining the effect of the bad option quality, we find a similar result to the symmetric discovery scenario. The stationary time increases with increasing $q_1$, in parallel with an increase in the value of consensus. However, when the system is unable to achieve consensus, the stationary time begins to decrease.

\begin{figure*}[t!]
    \centering
    \includegraphics[width=0.66\textwidth]{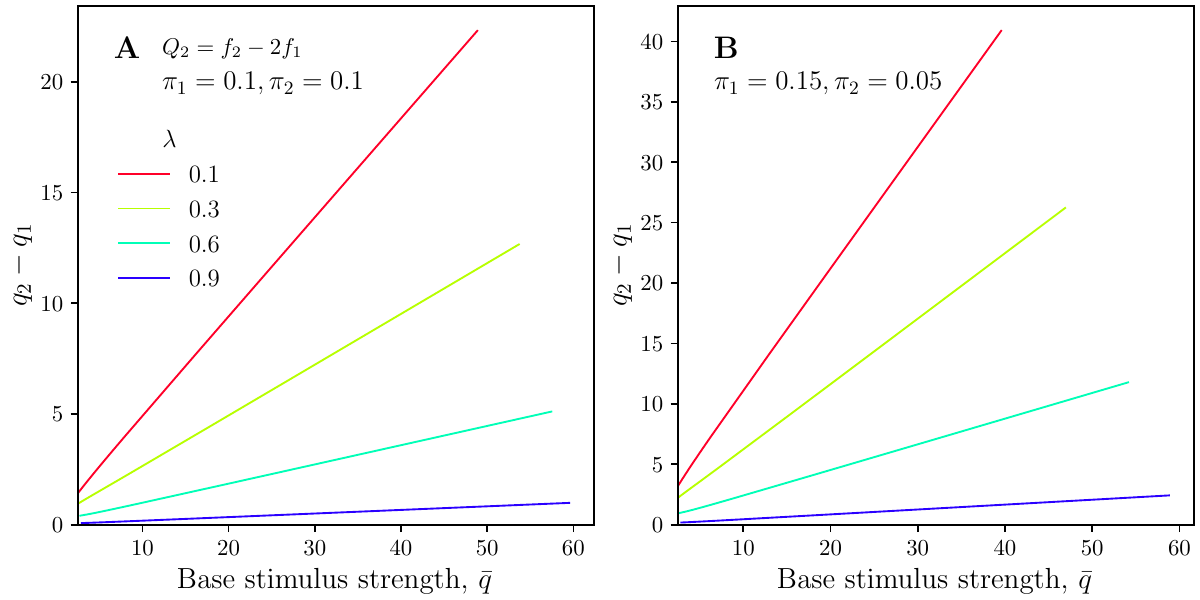}%
    \includegraphics[width=.33\textwidth]{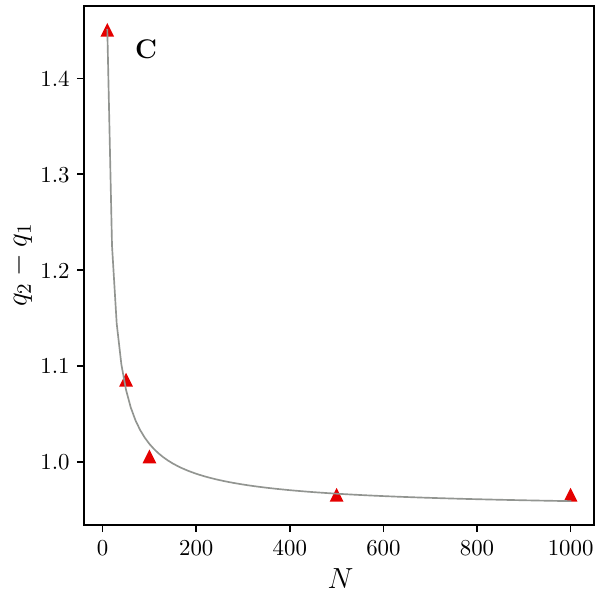}
    \caption{{\em Just noticeable quality difference} as a function of the base stimulus strength (\textbf{A}, \textbf{B}) and the system size (\textbf{C}).
    \textbf{A}: Symmetric discovery scenario with $\pi_1 = \pi_2 = 0.1$, \textbf{B}: Asymmetric discovery scenario with $\pi_1 = 0.15 > \pi_2 = 0.05$. Results obtained from analysis of the deterministic equations. \textbf{C}: Simulations for finite system sizes at a fixed $\lambda = 0.6$ and $\pi_{1,2} = 0.6$. The discernible quality difference decreases with the system size.}
    \label{fig:weber_law}
\end{figure*}

\subsection{Weber's Law of perception}\label{sec:webers_law}
Psychophysics explores how organisms perceive external stimuli, initially concentrating on human perception before broadening its scope to encompass other organisms across varying levels of biological complexity. There's a proposition that a swarm engaged in decision-making tasks can be seen as a superorganism~\cite{passino_swarm_2008}, exhibiting cognitive traits akin to those found in vertebrate brains based on neurons. Building on recent research into honeybee-inspired decision-making models~\cite{reina_psychophysical_2018, pais2013}, we aim to ascertain whether the LES model also reflects such characteristics.

We particularly investigate Weber's Law~\cite{fechner_elements_1948}, which posits that the smallest perceptible change between two stimuli (known as the just noticeable difference) is proportional to the intensity of the base stimulus. In our context, the stimuli correspond to the qualities of the two options under consideration in a binary decision problem, and the base stimulus intensity can be defined as the mean quality of the options, denoted as $\bar{q}$. According to Weber's Law, we anticipate observing a linear correlation between the difference in quality and the base stimulus strength ($\bar{q}$), expressed as $q_2 - q_1 = w \bar{q}$, where $w$ represents the Weber fraction.

To evaluate whether a swarm accurately discriminates between two stimuli, achieving consensus is necessary. We assess compliance with Weber's Law by keeping the quality of one option constant while varying the quality of the other until positive consensus is reached. In our experiment, we fixed the high-quality option while adjusting the inferior-quality option, but similar outcomes were obtained when reversing this setup. Figures~\ref{fig:weber_law}A and B confirm the linear correlation between the difference in quality and the stimulus strength across different combinations of interdependence and discovery probabilities. This analysis was conducted using the original consensus definition, Eq.~\eqref{eq:Qdefinition}, although similar results were obtained with alternative definitions discussed earlier. 

We tested the model agreement with Weber's Law in the two different discovery scenarios, symmetric ($\pi_1 = \pi_2$ Fig.~\ref{fig:weber_law}A) and asymmetric ($\pi_1 > \pi_2$, Fig.~\ref{fig:weber_law}B). While the discovery probabilities influence the actual quality differences discernible for a given level of interdependence, they do not affect the linear relationship with the base stimulus strength. A linear fit of the quality differences to the base quality, $q_2 - q_1 \sim w\bar{q}$, results in $R^2>0.99$ (we provide the fit parameters in Table I in the Supplementary Material). All these results suggest a general agreement with the Weber's Law.

Following the analysis in~\cite{reina_psychophysical_2018}, one can also relate the magnitude of finite-size fluctuations within the decision process with the random fluctuations typically observed in a discrimination process. These fluctuations can influence an organism's ability to accurately discriminate between two stimuli~\cite{thurstone_psychophysical_1927,stevens_psychophysical_1957}. It is widely recognized that collective decisions tend to improve when made by larger groups~\cite{Austen-Smith_Banks_1996, marshall2017_group_decision}.
As discussed in Sec.~\ref{sec:finite_size}, fluctuations diminish as the system size increases. Consequently, we anticipate that the system's ability to discriminate between two similar stimuli will improve with the system size, i.e. the just noticeable difference will decrease with $N$. In agreement with~\cite{reina_psychophysical_2018}, we observe this effect in our model, as illustrated in Fig.~\ref{fig:weber_law}C, which also exhibits a comparable exponential trend.

In the study by Reina et al.~\cite{reina_psychophysical_2018}, individual commitment transitions (i.e., discovery) are linked to site qualities, unlike in our model. Here, the quality-sensitivity influences system dynamics solely in the uncommitment and recruitment transitions. However, this difference does not appear to impact the model's adherence to Weber's Law. The independent discovery probabilities solely affect the specific value of quality differences that can be discerned. Figures~\ref{fig:weber_law} A and B also demonstrate that, for a given stimulus strength, the ability to discriminate decreases with $\lambda$. In other words, stronger social interaction enables the system to distinguish between more similar options, but this comes at the expense of a slower decision-making process, as previously noted in Figs.~\ref{fig:sym_Q_and_time} and \ref{fig:asym_pi_Q_lines_and_time}.

Contrary to our findings, Reina et al. in \cite{reina_psychophysical_2018} observe that the noticeable quality difference increases (while the decision time decreases) with the signaling ratio, which in their setup reflects the strength of social interactions relative to individual transitions. This disparity underscores the main distinction between the two models: the presence (or absence, in our case) of negative feedback in social interactions, specifically cross-inhibition. Introducing this mechanism in the decision process leads to quicker decisions, especially in situations involving similarly or equally valued options \cite{reina_model_2017}, albeit at the expense of reduced accuracy in the final outcome. These results emphasize not only the significance of social interactions but also their nature, as the presence or absence of cross-inhibition yields markedly different outcomes in scenarios initially featuring very similar problems.

\subsection{Finite Size Effects}\label{sec:finite_size}

\begin{figure}[t!]
    \centering
    \includegraphics[width=.8\columnwidth]{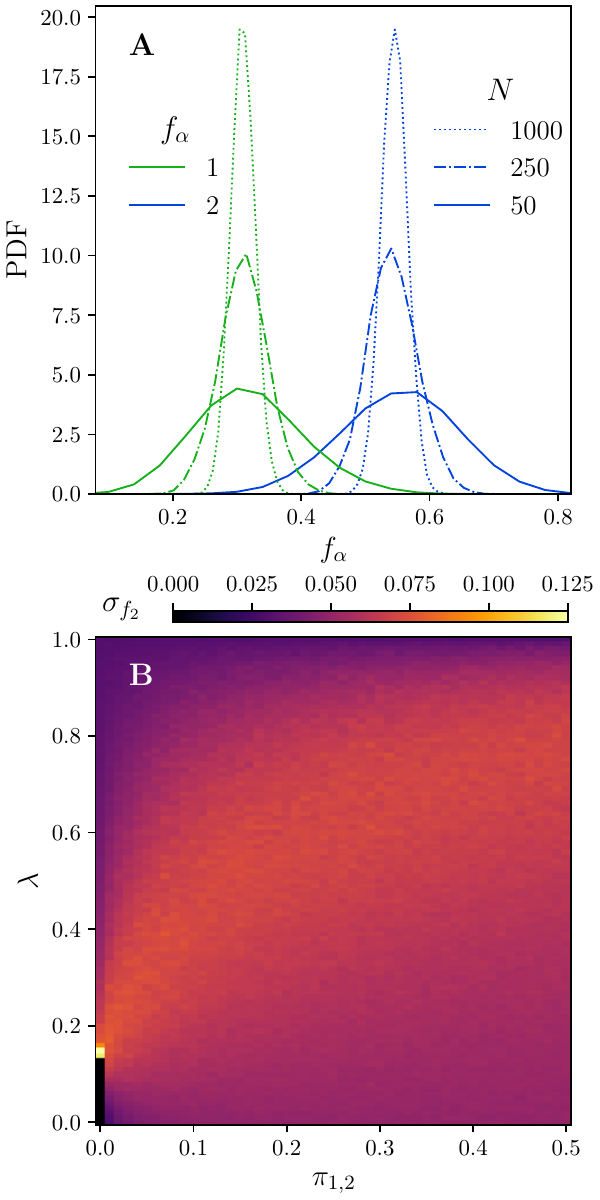}
    \caption{\textbf{A}: Probability density functions (PDFs) of the stationary site frequencies $f_1$ (green) and $f_2$ (blue) in simulations with different system sizes $N$. Other parameters are $\pi_1 = \pi_2 = 0.3$, $q_1 = 7$, $q_2 = 10$ and $\lambda = 0.3$. \textbf{B}: Variation of the standard deviation of the PDF for site 2, ($\sigma_{f_2}$), across the symmetric discovery parameter space, obtained from simulations with $N=100$ and $q_1 = 7$, $q_2 = 10$.}
    \label{fig:fluctuations}
\end{figure}

\begin{figure}
    \centering
    \includegraphics[width=.8\columnwidth]{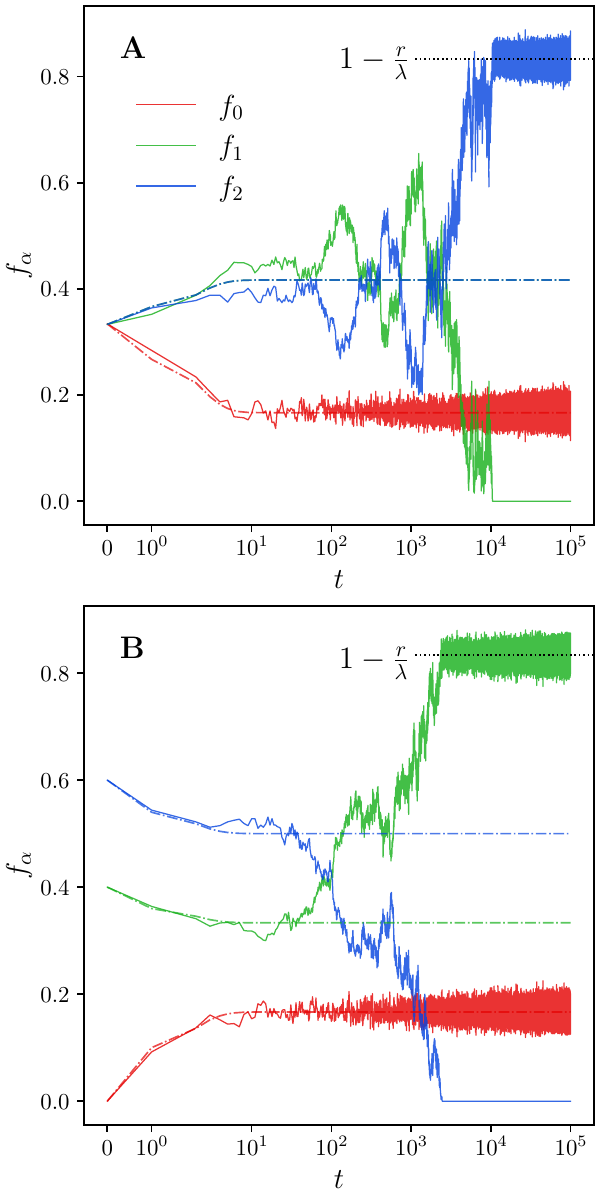}
    \caption{Temporal evolution of a single experimental realization in the fully interdependent limit ($\pi_\alpha = 0 \; \forall \alpha$) with equal qualities ($q_\alpha = 10 \; \forall \alpha$), starting from different initial conditions. \textbf{A}: $f_\alpha (t=0) \simeq (1/3, 1/3, 1/3)$ \textbf{B}: $f_\alpha (t=0) \simeq (0.0, 2/5, 3/5)$. Other parameters are $\lambda = 0.6$, $N = 1000$. The black dotted line indicates the value in which the \textit{winning} population stabilizes, $1 - r/\lambda$, while the other vanishes and $f_0 = r/\lambda$.}
    \label{fig:evos_pi0_q10}
\end{figure}

The deterministic solution analyzed in the previous subsections aligns perfectly with the results obtained by averaging the stationary state attained in mean-field simulations across all parameter space. However, stochastic simulations enable us to observe significant finite-size fluctuations. An example of such fluctuations for a specific choice of parameters is illustrated in Fig.~\ref{fig:fluctuations}A, where we show the probability density functions (PDFs) of stationary values of the dancing frequencies $f_1$ and $f_2$ obtained in stochastic simulations on fully connected systems of various sizes ($N = 50, \; 250, \; 1000$). The variance of the PDFs scales as $N^{-1}$ due to the central limit theorem. As a result, when the system size is relatively small, there is a notable overlap between the PDFs representing the optimal (site 2) and suboptimal (site 1) options. This indicates that a finite stochastic system may temporarily exhibit at least a simple consensus for the suboptimal (bad quality) option ($f_1 >f_2$). However, as the total population increases, the width of these distributions decreases, enhancing the robustness of the decision-making process. T. Galla further characterized these finite-size effects using the van Kampen expansion on the inverse system size \cite{Galla_2010}.

The interplay between independent exploration and social feedback also affects the magnitude of the fluctuations. In the same state space representation as in the previous sections, Fig.~\ref{fig:fluctuations}B illustrates the standard deviation of $f_2$ in the symmetric scenario. Excluding some extreme values near or at the fully interdependent limits, the fluctuations are most pronounced at intermediate levels of interdependence. 
Moreover, the region of high fluctuations appears to shift towards greater $\lambda$ on increasing $\pi_{1,2}$, 
akin to the consensus crossover lines discussed earlier.
 
The main rationale behind this observation lies in the intermediate region of the state space, where the competition between the two primary drivers of the system --independence and interdependence-- is most pronounced. 
When high levels of interdependence drive the system, 
the optimal site is easily favoured by the interplay of longer advertisement and high social feedback.
Conversely, when $\lambda$ is low, the stationary state is primarily determined by independent discoveries, with minimal or no influence from peer communication. However, in the central region, neither mechanism is dominant enough to outweigh the other, resulting in broader stationary distributions. Despite this broader distribution, consensus may still be established on average.

In the lower-left corner of Fig.~\ref{fig:fluctuations}B, we observe a null variance, followed by a sudden increase at higher values of $\lambda$. This corresponds to the region where the system ends up in a fully uncommitted state, occurring when $\pi_{1,2} = 0$ and $\lambda < r_2$ in an infinite system (recall Sec.~\ref{sec:fully_interdep_lim}). Although analytically the change in behavior is predicted to occur precisely at $\lambda = r_2$, we note that due to finite-size effects, this behavior extends slightly beyond this threshold. 
At $\lambda \gtrsim r_2$, fluctuations suddenly increase as most of the simulation results converge around the predicted stable solution, $f_0 = r_2/\lambda$, while, due to finite-size effects, some results still show the system stabilizing at the fully uncommitted state. As the system size increases, the transition becomes sharper around the threshold $\lambda = r_2$. 

The stationary results and phase diagrams for the case where $\pi_\alpha \neq 0$ are independent of the initial conditions, as the system consistently evolves towards a unique stable fixed point. However, initial conditions do affect the deliberation time required to reach this stationary state. For this reason, we have chosen to analyze the system's evolution starting from a fully uncommitted initial state, as it closely resembles a natural decision-making scenario.

 At or near the fully interdependent limit ($\pi_\alpha = 0$), the effects of initial conditions become more pronounced. As the system approaches this limit, it undergoes critical slowing down due to the phase transition at $\lambda = r_2$ when $\pi_{1,2} = 0$ (see App~\ref{sec:appendixA}). In such cases, the bias introduced by initial conditions can significantly extend the decision time. Furthermore, in finite systems, the interplay between fluctuations and unfavorable initial conditions may lead the system to deviate from the stable fixed point—where the superior option is typically selected-for extended periods, particularly when $q_1 \rightarrow q_2$. In the exact limit $\pi_{1,2} = 0$, the system could even select the lower-quality option, with initial conditions introducing a strong bias that can significantly influence the consensus outcome.

Lastly, let's consider the scenario where the site qualities are equal (and their discovery probabilities are also equal) in the fully interdependent limit ($\pi_\alpha = 0$). As we've discussed in the analytical solution of the model, detailed in Sec.~\ref{sec:fully_interdep_lim}, the system undergoes a phase transition between a fully uncommitted absorbing state and an active state with finite fractions of bees advertising the available sites at $\lambda = r$. We further investigate this non-equilibrium phase transition in the following section. Here, we briefly emphasize the influence of finite-size effects under this limit. 
While the system may temporarily fluctuate around the stationary state predicted by the deterministic equations, finite-size effects ultimately break this tendency, and the system converges towards a consensus for only one of the options, while the other option disappears. In Figs.~\ref{fig:evos_pi0_q10}, we illustrate this phenomenon, 
by representing the temporal evolution of $f_0$, $f_1$ and $f_2$ for a binary decision problem starting from two different initial conditions.
Continuous lines represent the results obtained in stochastic simulations of a finite system consisting of $N=1000$ bees, while dashed lines represent the integration of deterministic equations Eqs.~\eqref{eq:model_time_evo}, representing the behavior of an infinite system. We observe that fluctuations introduced by finite-size effects ultimately break the symmetry between the two available options. This can occur even when one option initially appears to be favored. The initial difference between options will also affect the likelihood of observing this switch.

In real-world applications, whether in ecological systems or synthetic ones like robot swarms, these fluctuations are always expected to be relevant because most of these systems are finite. An important conclusion can be drawn from these results. In ecological systems, the ability to regulate the dissemination of information is crucial for efficient decision-making. For instance, when options have comparable qualities, it may be beneficial for the group to converge quickly towards a consensus to avoid prolonged indecision. In such cases, certain members within the group may play a pivotal role by ceasing exploration or transmitting stop exploration signals to others~\cite{seeley_stop_2012, reina_cross_inhibition_2023, marchpons2024arXiv_nonlinCI}. These signals would effectively reduce the parameter $\pi$ to near zero, signaling to other group members to cease exploration and focus on promoting consensus for one of the options. As we will analyze in the following section, reducing this parameter brings the system closer to the state where maximal consensus is attained. 
Thus, this adaptive behavior allows the group to navigate decision-making scenarios with similar options more effectively, potentially avoiding resource wastage. However, deterministic modeling approaches, may struggle to capture the dynamics of decision-making processes in real finite systems in similar circumstances.

\subsection{Absorbing phase transition and finite size effects in the fully interdependent limit}\label{sec:phase_trans_cp}

The analytical analysis of the model in Section~\ref{sec:equal_qualities_sol} reveals that in the fully interdependent limit ($\pi_{\alpha} = 0$ for all $\alpha$), the system undergoes a phase transition between two states: an absorbing uncommitted state and a partially-active state. In this limit, the system relies entirely on imitation. Thus, if the imitation parameter $\lambda$ exceeds a critical value $\lambda_c$, imitation alone is sufficient to sustain a steady state with a finite committed population of bees, denoted as $\rho = \sum_{\alpha = 1}^{\alpha = k} f_{\alpha}=1-f_0$, at least for the highest quality option. However, below $\lambda_c$, no bees actively advertise any site, and the system remains locked in a state where $f_0=1$.

Moreover, when qualities are equal for all sites, this problem can be exactly mapped to the well-known contact process, which exhibits the same non-equilibrium critical behavior. In this framework, the parameter $\rho=1-f_0$, representing the fraction of agents actively promoting available sites, acts as the order parameter of the transition. Therefore, we anticipate that sufficiently close to the critical threshold $\lambda_c$,
\begin{equation}
    \rho \sim (\lambda-\lambda_c)^\beta
\end{equation}
with $\beta$ a dimension-dependent critical exponent, which in the mean field approximation is exactly equal to $\beta = 1$. 

\begin{figure}[t!]
    \centering
    \includegraphics[width=.8\columnwidth]{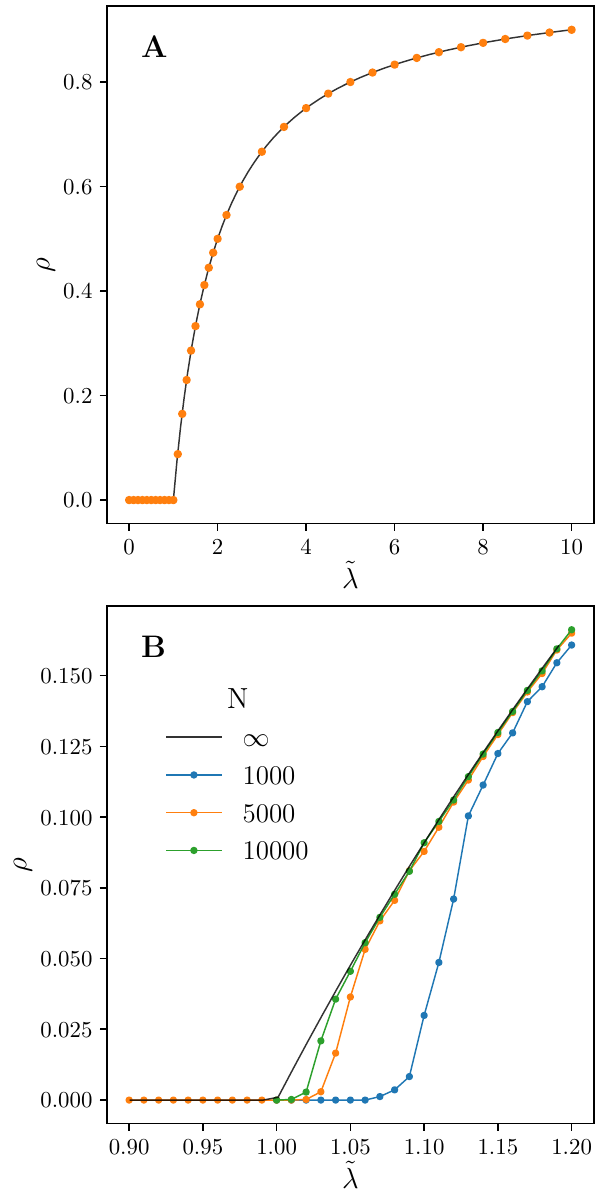}
    \caption{Absorbing phase transition in the fully interdependent limit ($\pi_\alpha=0, ~ \forall \alpha$) with equal qualities ($q_\alpha=10, ~ \forall \alpha$) for a fully-connected system.
    \textbf{A}: Order parameter $\rho=1-f_0$ vs. $\Tilde{\lambda}=\lambda /\lambda_c$.
    \textbf{B}: Zoom around the phase transition happening at $\Tilde{\lambda}_c = 1$.
    }
    \label{fig:phase_transition_lc_meanfield}
\end{figure}

\begin{figure}[t!]
    \centering
    \includegraphics[width=0.47\textwidth]{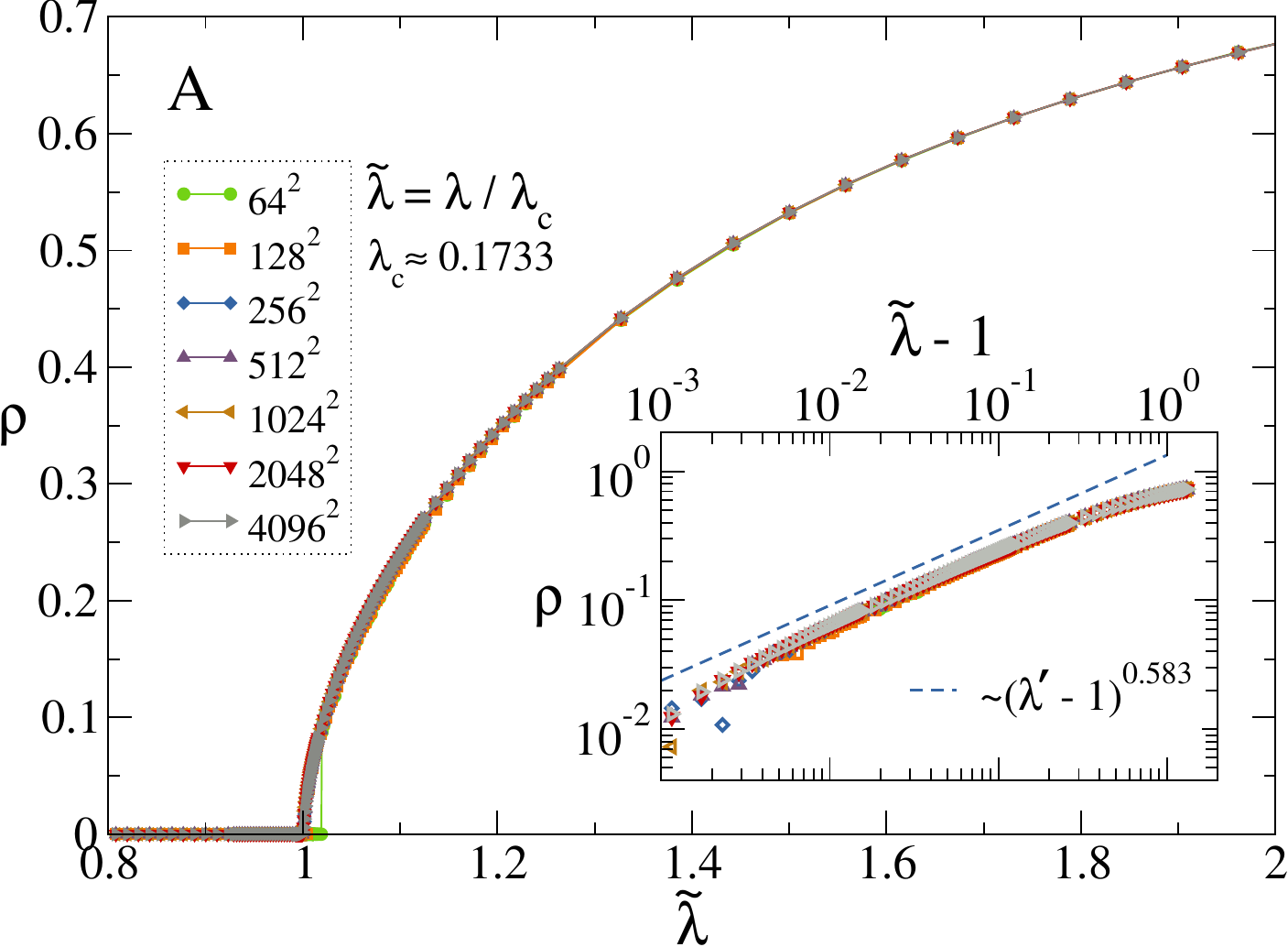}\\
    \mbox{ } \\
    \includegraphics[width=0.49\textwidth]{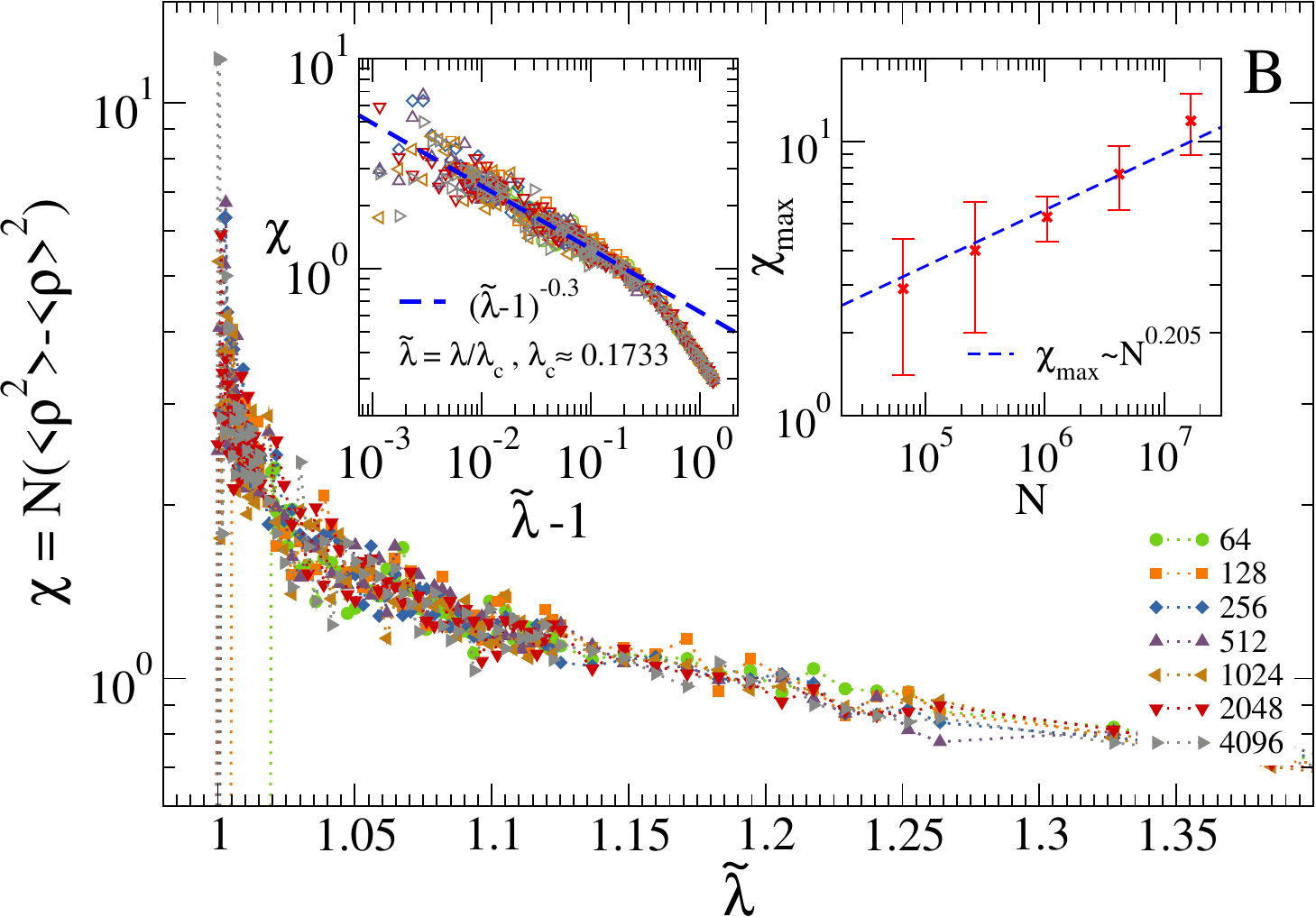}
    \caption{
    Phase transition in the fully interdependent limit ($\pi_\alpha=0 ~, ~ \forall \alpha$) with equal qualities ($q_\alpha=10, ~ \forall \alpha$) for stochastic simulations of a binary decision process on a regular square lattice.
    \textbf{A}: Order parameter $\rho=1-f_0$ vs. $\tilde{\lambda}=\lambda /\lambda_c$ for different system sizes. The inset shows $\rho$ as a function of $\tilde{\lambda} -1$.
    A power-law $\rho \sim (\tilde{\lambda} -1)^\beta$ is displayed with $\beta=0.583$ 
    (blue dashed-line) to show consistency with DP exponents. 
    If fitted, $\beta \simeq 0.58 \pm 0.02$, where the error comes from the variability when changing the fitting region and the uncertainty in the determination of $\lambda_c$.
    \textbf{B}: Susceptibility $\chi = N \left(\left<\rho^2\right>-\left<\rho\right>^2\right)$ vs. $\tilde{\lambda}=\lambda /\lambda_c$ for different system sizes. 
    Insets: 
    (Left) Approaching the critical point from above, the susceptibility scales in a way compatible with $(\lambda/\lambda_c -1)^{-0.3}$. 
    (Right) Finite-size scaling. At the critical threshold $\lambda_c \simeq 0.1733$, the susceptibility scales in agreement to $\sim N^{0.205}$. Scaling laws show the expected DP critical exponents rather than fitting directly the data.
    }
    \label{fig:phase_transition_lc_lattice}
\end{figure}

Indeed, we can rewrite our deterministic differential equations (Eq.~\eqref{eq:model_time_evo}) by setting $\pi_{\alpha} = 0$ and $r_{\alpha} = r$ for all $\alpha$:

\begin{equation*}
    \langle \dot{f}_{\alpha} \rangle = \lambda \langle f_0 \rangle \langle f_{\alpha} \rangle - r \langle f_{\alpha} \rangle \qquad \forall \alpha \text{.}
\end{equation*}

\noindent Summing over all sites, one gets the mean field rate equation for the density of active sites,
\begin{equation}
    \frac{d\rho}{dt} 
    = \lambda (1 - \rho) \rho - r \rho \text{.}
\end{equation}
Rescaling by $r$, to define $\tilde{\lambda} = \lambda/r$ and $\tau = rt$, and rearranging, we arrive at the well-known mean-field equation for the contact process:
\begin{equation}
    \frac{d\rho}{d\tau} = (\tilde{\lambda}-1)\rho - \tilde{\lambda} \rho^2 \text{.}
\end{equation}
This equation has two stationary points: $\rho^* = 0$ and $\rho^* = 1-\tilde{\lambda}^{-1}$, with the mean-field critical point located at $\tilde{\lambda}_c = 1$. As detailed in Section~\ref{sec:equal_qualities_sol}, the absorbing state is stable if $\tilde{\lambda} < 1$, while the active state is stable when $\tilde{\lambda} > 1$.
Fig.~\ref{fig:phase_transition_lc_meanfield} shows $\rho$ vs $\Tilde{\lambda}$ for
simulations of the fully-connected model alongside the analytical deterministic solution. As anticipated, the analytical solution exhibits the transition at $\tilde{\lambda}_c = 1$. However, due to finite size effects, simulations show deviations from this result, with larger threshold values than expected for an infinite system, $\tilde{\lambda}_c(N) > \tilde{\lambda}_c$.

The self-discovery probabilities $\pi$ play the role of an external field, as with non-null $\pi$ values, an inactive site can spontaneously become active. Similar to the contact process, the introduction of this external field disrupts the absorbing state and consequently the transition itself, leading the system away from criticality. However, for sufficiently small values of $\pi$, the system still obeys certain scaling laws, as discussed in~\cite{Hinrichsen2000}. We have also confirmed that, for instance, the order parameter in stationary conditions is compatible with the scaling law $\rho \sim \pi^{\beta/\sigma} \sim \pi^{1/2}$, with mean-field exponent values $\beta = 1$ and $\sigma = 2$.

The contact process falls within the directed percolation (DP) universality class. Therefore, beyond mean field, the critical exponents of the uncommitted absorbing phase transition should match those of directed percolation, which depend on the system spatial dimension. For completeness, we also analyze the phase transition on a regular 2-dimensional square lattice with nearest-neighbors interactions. Further details about simulations of the LES model in this geometry are provided in Appendix \ref{sec:appendixE}.
In Fig.~\ref{fig:phase_transition_lc_lattice}A, we show the density of active sites around 
the transition for simulations of different system sizes $N$ and $q_1=q_2=10$ (or equivalently $r=0.1$).

A finite-size scaling analysis enables us to identify the critical interdependence parameter 
value $\lambda_c \simeq 0.1733$. 
Beyond this threshold, we note a close correspondence with an order parameter scaling 
behavior characterized by the exponent $\beta \simeq 0.583$, as expected for DP in two dimensions~\cite{Hinrichsen2000}. 
Furthermore, as in other phase transitions, one can define a susceptibility for the 
decision-making problem in this limit as 
$\chi = N \left(\left<\rho^2\right>-\left<\rho\right>^2\right)$, 
which exhibits a peak at the pseudo-critical threshold $\lambda_c(N)$. 
For directed percolation, it is expected to scale as
$\chi \sim L^{(d\nu_\perp - 2\beta)/\nu_\perp}$~\cite{JensenPRE1993,lubeck_universal_2004}, 
where for $d=2$, $\beta=0.583$ and $\nu_\perp =0.733$. 
Therefore, $\chi \sim L^{0.409} \sim N^{0.205}$. 
The scaling properties of this quantity in our simulations on the square lattice 
are depicted in Fig.~\ref{fig:phase_transition_lc_lattice}B. 
As expected, the susceptibility peaks at $\lambda_c(N)$ for different system sizes $N$. 
At the critical threshold $\lambda_c \simeq 0.1733$, it scales in a way compatible with 
$\chi \sim N^{0.205}$, the expected scaling behavior.
Additionally, on approaching the critical point from the active phase, the susceptibility 
diverges roughly as $\chi\sim (\lambda/\lambda_c -1)^{-\gamma'}$, 
with $\gamma' = d\nu_\perp - 2\beta = 0.3$. 
The measured exponent value is also compatible with a DP exponent in $d=2$~\cite{lubeck_universal_2004}. Therefore, our results consistently indicate that in the fully interdependent limit, our model reduces to a DP-like process, with a dimension-dependent non-equilibrium critical phase transition undergoing the building of consensus in the swarm.

Non-equilibrium phase transitions are commonly observed in biological 
systems~\cite{MAMunoz_review2018}. 
These transitions, which occur far from thermodynamic equilibrium, play a crucial role in 
shaping the collective behavior of living organisms. 
As a matter of fact, we observe that this model system achieves optimal consensus 
at the onset of critical behavior, when approaching the fully interdependent limit $\pi_{1,2} \rightarrow 0$ (or $\lambda \rightarrow 1$), identifying this as the optimal operating point~\cite{mora_bialek_2011, Chialvo2020}. 
Biological systems often incorporate adaptive mechanisms, such as quorum-dependent stop
signals~\cite{seeley_stop_2012}, which can modulate interdependent activity or discourage independent exploration. 
These mechanisms allow them to eventually operate near the critical point, once sufficient information has been gathered from the environment. 
The adaptive ability to operate around a critical point is particularly important for 
collective decisions involving equivalent options (i.e., options of the same quality and 
discovery probability), as symmetry can only be broken by critical fluctuations around 
this optimal operating point. 
While our current model lacks an explicit mechanism to adjust the self-discovery parameter, 
our parametric study reveals that operating near this critical point yields maximum consensus 
(see Figs.~\ref{fig:sym_Q_and_time}A,B). 
Developing model extensions to incorporate these features would be an interesting direction for future research.

\subsection{Multiple sites}\label{sec:more_sites}

\begin{figure}[t!]
    \centering
    \includegraphics[width=.8\columnwidth]{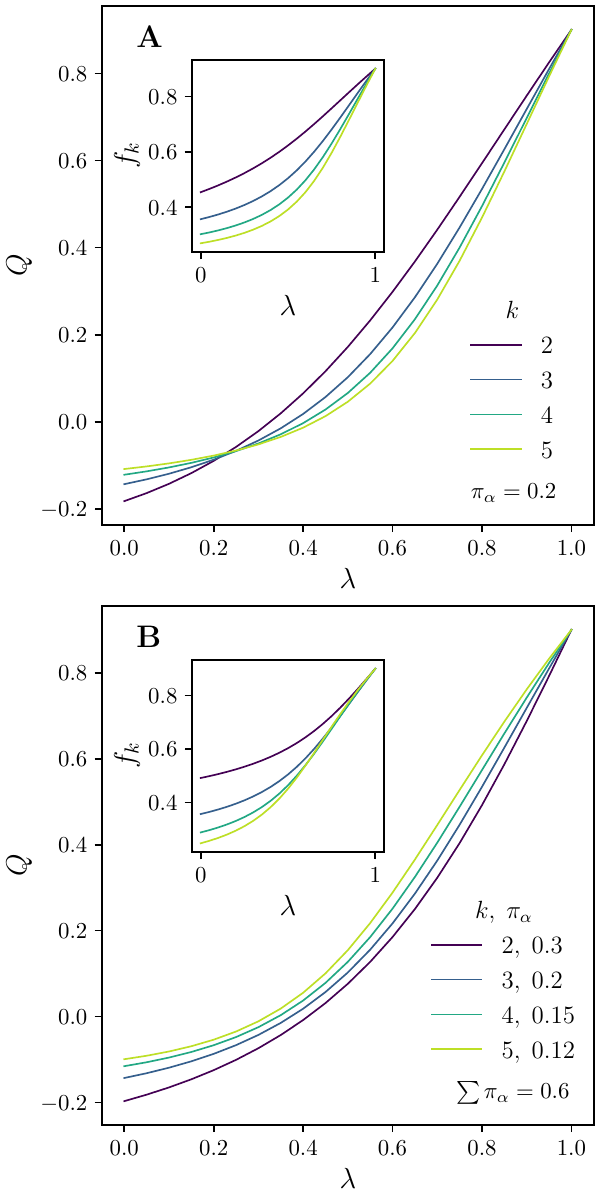}
    \caption{Consensus parameter $Q$ and dance frequency for the best site $f_{k}$ (shown in the inset) as a function of the interdependence for different numbers of options, $k$. The qualities of the sites are defined as follows: $q_k = 10$, $q_{k-1}=7$, and $q_{k-i}=q_{k-i+1}-1$ for $2\leq i \leq k-2$ for options below the top two. \textbf{A}: Each site has a self-discovery probability $\pi_{\alpha}=0.2$. \textbf{B}: Self-discovery probabilities are equal and their sum remains constant, $\sum_{\alpha=1}^k\pi_{\alpha}=0.6$, so as the number of sites increases, $\pi_{\alpha}=0.6/k$ decreases.}
    \label{fig:more_sites}
\end{figure}

Up to now, our attention has been directed towards a binary decision problem, where selecting the best option is imperative.
Many collective decision making models have been concerned in generalizing their dynamics to multiple option scenarios~\cite{StarniniJSTAT2012,reina_model_2017, VazquezPRE2019, RamirezPRE2022}.
Particularly, previous research on this model, utilizing different parameter setups, examined a specific scenario involving five options. It's worth noting that increasing the number of available options doesn't qualitatively alter the model's behavior. In a general $k$-site scenario, we still observe the same interplay between independent discovery and social feedback. At high $\lambda$ values, the system effectively identifies the best site, while at low $\lambda$ values, it settles into a multi-opinion state where each site is occupied in proportion to its quality and discovery probability. Therefore, in a broader best-of-N scenario, the specific dance frequency values and consensus transitions will rely on the $\pi_{\alpha}$ and $q_{\alpha}$ parameters defining each unique scenario. However, the overall behavioral trend of the system remains consistent across different setups.

In Fig.~\ref{fig:more_sites} we represent the consensus parameter together with the dance frequency for the best-site $f_k$ (depicted in the inset) for different number of available sites $k$. We consider two different symmetric scenarios: 
one where each site has the same self-discovery probability (independently of the number of sites $k$), and another where the total sum of self-discovery probabilities remains constant, i.e., $\sum_{\alpha=1}^k\pi_{\alpha}=0.6$, and consequently each time a new site is added the overall value of $\pi_{\alpha}=0.6/k$ decreases. 
In both scenarios, site qualities are different and ordered following the sequence $q_1 < q_2 < ... < q_k$. While there is a slight quantitative change with increasing $k$, the system's behavioral trend observed in the binary problem remains consistent. 

As the number of available options with the same discovery probability increases, the system is exposed to a greater amount of external information. Consequently, slightly higher values of $\lambda$ are required to achieve the consensus crossover, as evidenced by the rightward shift of the consensus curves in Fig.~\ref{fig:more_sites}A around $Q = 0$.
Contrarily, when the overall level of external information ($\sum \pi_{\alpha}$) remains constant but the individual discovery probabilities $\pi_{\alpha}$ decrease as the number of options increases, as shown in  Fig.~\ref{fig:more_sites}B - the system achieves consensus more readily as the number of options increases, a result that may appear counter-intuitive compared to other studies on the best-of-N scenario with different models~\cite{reina_model_2017}. However, this outcome is a direct consequence of the decision mechanisms embedded in our model: increasing $\lambda$ strengthens the positive feedback on populations associated with the best quality, leading to the rapid elimination of options with lower qualities from consideration. Consequently, the decision process becomes nearly binary between the top two options. Decreasing the overall values of the discovery probabilities simplifies the discussion process further, facilitating consensus. 
Supplementary Fig. 1 in~\cite{suppmat} illustrates all dance frequencies $f_{\alpha}$ in the reported scenarios, demonstrating the consistently decreasing trends of the remaining populations in favor of the highest-quality option as $\lambda$ increases.

Other scenarios, such as a generally asymmetric discovery scenario (where $\pi_1 > \pi_2 > ... > \pi_k$) or equivalent inferior options ($q_1 = ... = q_{k-1}$), exhibit similar trends to the discussed symmetric discovery scenario. Naturally, the specific values of $f_k$ or the consensus may vary, as it becomes more challenging for the swarm to achieve consensus under these conditions. However, the interplay between interdependence, site qualities, and discovery probabilities remains unchanged. 
In Supplementary Fig. 2 in~\cite{suppmat} we present the results obtained in these scenarios to emphasize the qualitative similarities discussed earlier.

\section{Conclusions}\label{sec:conclusions}

We have performed an analytical study of an agent-based model inspired by the collective behavior of honeybees~\cite{list_independence_2008}, following the mathematical framework outlined in~\cite{Galla_2010}. This model incorporates key features such as the balance between independent exploration and social interdependence, as well as the sensitivity of agents to option quality. Using analytical reductions of this agent-based model, we extensively analyzed its behavior in the space of relevant parameters. Our goal was to investigate whether the system achieves consensus for the best available option when balancing information acquisition and social interactions. Furthermore, we supplemented this analysis with simulations to report the time required for the system to converge to its stationary state.

We have explored two scenarios: one where options are equally likely to be discovered, and another where the worst quality option (among two) is more likely to be discovered. In both cases, when the system prioritizes interdependence, the established consensus is stronger.
This highlights the role of social interactions in achieving a higher accuracy in the final decision, consistent with similar honeybee-inspired models~\cite{reina_desing_pattern,reina_model_2017,gray_multiagent_2018}. Interdependence plays a critical role, especially when the system deals with high values of the self-discovery probabilities or when the superior option faces a disadvantage in terms of discovery likelihood. In such scenarios, interdependence emerges as a crucial {\em noise}-reducing mechanism, allowing the system to maintain a high degree of coherence even when individuals are more prone to random changes in their state.

However, escalating social feedback diminishes the extent of individual exploration, leading to a prolonged time required to reach the stationary state. This observation underscores a trade-off between speed and accuracy, indicating that a moderate level of social interactions is most beneficial when a system seeks to balance consensus accuracy and convergence time~\cite{passino2006}. This finding aligns with previous research investigating the emergence of a speed-accuracy trade-off in opinion dynamics models~\cite{valentini2016, reina_voter_2023}. In those studies, interactions governed by a majority rule revealed a trade-off between decision accuracy and the size of the interaction group.

In our binary decision task analysis, we have further proved the model's compliance with Weber's Law of psychological perception, which suggests that the smallest noticeable change between two stimuli is proportional to the intensity of the base stimulus~\cite{fechner_elements_1948}. This relation was first studied in individual organisms, but recently efforts have been devoted to study swarms faced with perception tasks as \textit{superorganisms}, capable of obeying the same laws~\cite{passino_swarm_2008,reina_psychophysical_2018}. This assessment was conducted in both symmetric and asymmetric discovery scenarios. While the discovery probabilities do impact the discernible quality differences for a given level of interdependence, they do not alter the linear relationship with the base stimulus strength. Thus, our findings indicate a broad concordance with Weber's Law. In this context, we have also confirmed that the system's capability to distinguish between two similar stimuli enhances with an increase in system size.

When confronted with a decision problem between two equal-quality options, the deterministic solution of the LES model predicts a system unable to break the symmetry and achieve consensus for one option. In such situations, the introduction of more complex higher-order interactions, such as cross-inhibition, becomes necessary. This need for higher-order interactions has been observed in both on-field honeybee experiments~\cite{seeley_stop_2012}, computational models~\cite{reina_desing_pattern,gray_multiagent_2018,reina_cross_inhibition_2023}, and robot swarm experiments~\cite{reina_2016_decentdecision,zakir_robot_2022,talamali_improving_2019,reina_cross_inhibition_2023}.
However, finite-size fluctuations can readily break the symmetry between the options if the system is driven around the so-called "fully interdependent" limit, where individual exploration behavior plays a minimal role. This finding suggests a simple adaptive behavior of the model: reducing the strength of the self-discovery probabilities as the decision process progresses, especially in scenarios where options are advertised similarly. By doing so, the system can break the symmetry even between identical options. However, it's important to note that while this mechanism allows for simple symmetry breaking, it makes the system less flexible when it needs to adapt in changing environments~\cite{talamali2021_less_more, aust_hidden_2022}.

Delving into the analysis of the fully interdependent limit, we have discovered an interesting correspondence with the well-known contact process~\cite{Hinrichsen2000}. 
In this limit, where the discovery probabilities are negligible, the dynamics are solely governed by interactions. By adjusting the interdependence, which represents the strength of these interactions, the system undergoes a non-equilibrium phase transition. Below a critical value $\tilde{\lambda}_c$, the system settles into an inactive, absorbing state—referred to as the uncommitted state in opinion dynamics terminology. Conversely, above the critical value, the system can maintain a non-negligible proportion of active or committed population for either of the two options, particularly when they are of equal quality. We have derived the model equations, which exactly map to the mean-field contact process equations. In addition, making use of $d=2$ lattice simulations, we have explored the critical behavior and some of the critical exponents surrounding the phase transition. Our findings confirm that the critical exponents of the order parameter and susceptibility consistently align with those reported for the contact process in $d=2$, which falls within the directed percolation universality class. 
Our study further reveals that the system achieves maximum consensus near this critical point. This finding aligns with the ongoing debate on how biological and bio-inspired systems operate near criticality~\cite{mora_bialek_2011, MAMunoz_review2018, Chialvo2020}. It reinforces the idea that adaptive mechanisms, such as inhibitory signals, may play a crucial role in reducing independent information gathering, allowing the system to rely more on intercommunication to reach maximum consensus. These mechanisms can guide the system to operate near this critical point, once enough environmental information has been pooled, optimizing decision-making efficiency.

While our primary focus has been on a binary decision problem, we have also evaluated the model's robustness when expanding the number of available options. Unlike other studies~\cite{reina_model_2017}, we observe that increasing the number of sites consistently improves the swarm performance, or accuracy, when making decisions among non-equivalent options. As discussed in~\cite{reina_model_2017}, the main distinction between our approaches lies in the absence or inclusion of negative social feedback, such as cross-inhibition between populations representing different options. Consequently, the effect of increasing social interactions, or signaling, has varying impacts on the system dynamics.

In conclusion, our study of a simple honeybee-inspired collective decision-making model reveals the intricate interplay between individual exploration and social interactions in shaping consensus and decision outcomes. We have demonstrated how the balance of these factors influences the system's ability to discriminate between options, and to achieve strong enough consensus. The role of finite-size or critical fluctuations becomes particularly relevant in decision-making processes of adaptive systems when available alternatives are very similar. Indeed, these fluctuations play a crucial role in breaking deadlocks or abandoning fully uncommitted states, thus avoiding potentially dangerous situations in ecological systems. Our findings contribute to a deeper understanding of collective behavior in biological systems and provide insights that may inform the design and optimization of decision-making algorithms in artificial systems.

\begin{acknowledgments} 
We acknowledge financial support from the Spanish MCIN/AEI/10.13039/501100011033, 
through projects PID2019-106290GB-C21, PID2019-106290GB-C22, PID2022-137505NB-C21 
and PID2022137505NB-C22.
D.M. acknowledges support from the fellowship FPI-UPC2022, granted by 
Universitat Polit\`ecnica de Catalunya.
E.E.F. acknowledges support from the Maria Zambrano program of the 
Spanish Ministry of Universities through the University of Barcelona
and PIP 2021-2023 CONICET Project Nº 0757.

\end{acknowledgments}

\appendix
\renewcommand{\thefigure}{A\arabic{figure}}
\setcounter{figure}{0}

\section{Appendix}

\subsection{Linear Stability Analysis on the general solution}\label{sec:appendixA}
Studying the effects of perturbations on the general deterministic solution up to linear order leads to a square-$k$ matrix with coefficients:
\begin{eqnarray*}
    & & a_{\alpha \alpha} = \lambda (f_0^* - f_{\alpha}^*) - (1-\lambda)\pi_{\alpha} - r_{\alpha} \\
    & & a_{\alpha \beta} = - (1 - \lambda) \pi_{\alpha} - \lambda f_{\alpha}^*
\end{eqnarray*}
\noindent where $\alpha=1,...k$. The stability of any solution can be verified by obtaining the matrix eigenvalues numerically, once the actual solution $(f_0^*, f_1^*, ..., f_k^*)$ is known.

\begin{figure}[t!]
    \centering
    \includegraphics[width=.7\columnwidth]{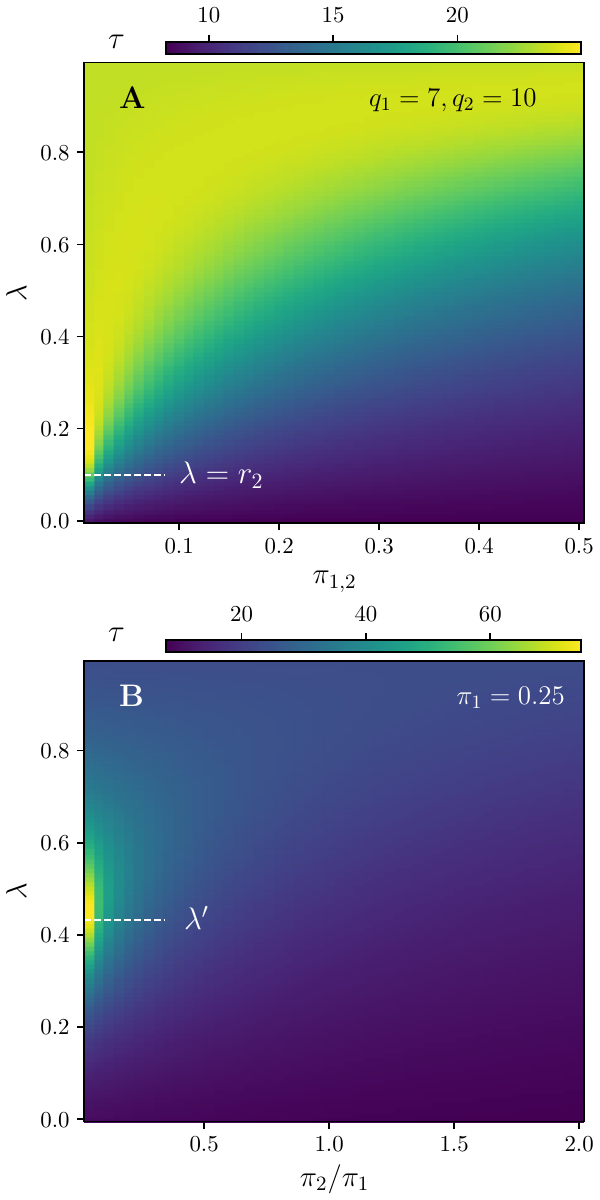}
    \caption{Characteristic relaxation times $\tau$ of linear perturbations obtained from the LSA eigenvalues. \textbf{A}: Symmetric discovery scenario. \textbf{B}: Asymmetric discovery scenario with $\pi_1 = 0.25$. In both scenarios, the qualities are $(q_1, q_2) = (7, 10)$. White dashed lines indicate the transitions occurring at the fully interdependent limit $\pi_{\alpha}=0$ (null-$\pi_k$ limit) for the symmetric (asymmetric) scenario. For the exact value of $\lambda' \simeq 0.43$, refer to Eq.~\eqref{eq:null_pi2_lambda_crit}.}
    \label{fig:char_times}
\end{figure}

The eigenvalues obtained from the numerical analysis of this matrix can be used to compute the relaxation time of a perturbation in the stationary state, denoted as $\tau$. This represents the characteristic time for the system to return to the state $(f_0^*,f_1^*,...,f_k^*)$ after a small perturbation $f_{\alpha}'$ is introduced in any value of $f_{\alpha}^*$ (note that these perturbations must satisfy that $\sum_{\alpha = 0}^{k} f_{\alpha}' = 0$). Specifically, the relaxation time is determined by the smallest eigenvalue, $\tau^{-1} = \text{min}(\theta_\alpha)$. In Fig.~\ref{fig:char_times}, we illustrate the characteristic relaxation times for a fixed set of qualities in both the symmetric and asymmetric discovery scenarios.

We observe that in the limit of small $\pi$, near the transition occurring at the fully interdependent limit (marked with a white-dashed line at the appropriate value of $\lambda$ in both scenarios), the relaxation times increase sharply. This suggests the onset of a critical slowing down of the system dynamics around these points.
Finally, it's worth noting the similarity between these relaxation times and the stationary times discussed in the main text (see sections~\ref{sec:sym_disc} and \ref{sec:asym_disc}, and Figs.~\ref{fig:sym_Q_and_time}C and \ref{fig:asym_pi_Q_lines_and_time}D). The same mechanisms that govern the convergence time to reach the stationary state also play a role when the system is relaxing from a small perturbation. However, in simulations of the asymmetric discovery scenario, the increase in time when $\pi_2 \rightarrow 0$ occurs at slightly larger values of $\lambda$, above the transition at $\lambda'$, and while it slightly decreases with increasing $\lambda$, it does not return to the level prior to the transition, unlike the relaxation time of linear perturbations.

\begin{figure*}[t!]
    \centering
    \includegraphics[width=.97\textwidth]{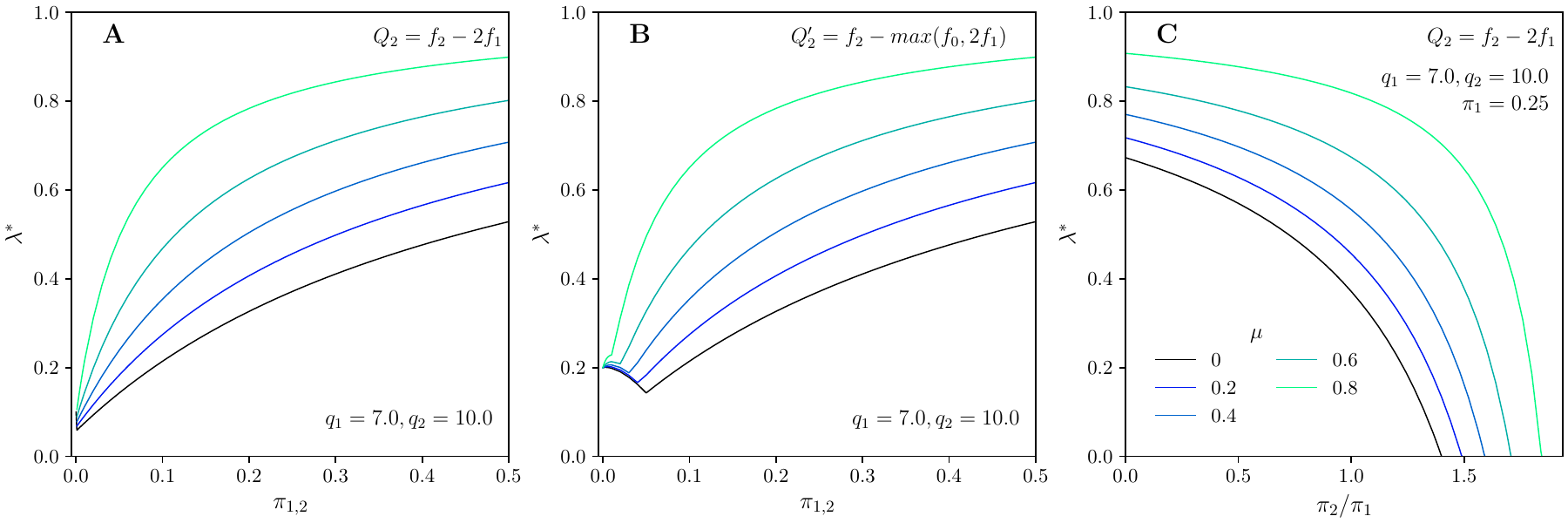}
    \caption{Consensus crossover lines for different values of the independent quality assessment parameter $\mu$. \textbf{A}: Consensus $Q_2$ in the symmetric discovery scenario. \textbf{B}: Consensus $Q_2'$ in the symmetric discovery scenario. \textbf{C}: Consensus $Q_2$ in the asymmetric discovery scenario with $\pi_1 = 0.25$ and varying $\pi_2$. The quality parameters are $q_1 = 7$, $q_2 = 10$.}
    \label{fig:Qlines_var_mu}
\end{figure*}

\subsection{Linear Stability Analysis on the Fully Interdependent limit}\label{sec:appendixB}

In the fully interdependent limit ($\pi = 0$), we encounter $k+1$ feasible fixed point solutions (see Eqs. (\ref{eq:f0_pi_0})). To ascertain their stability, we conduct a linear stability analysis. Our analysis reveals that the best-site winning solution $f_k^*$ yields all negative eigenvalues, indicating its stability when $\lambda > r_k$. Specifically, the resulting eigenvalues are as follows:
\begin{eqnarray*}
   & & \theta_m = r_k - r_m < 0 \quad m = 1,...,k-1 \\ 
   & & \theta_k = r_k - \lambda < 0 \quad \text{if} \quad \lambda > r_k
\end{eqnarray*}
If $\lambda < r_k$, this fixed point becomes unstable, and the stable solution is the absorbing fixed point $f_0^*=1$. The eigenvalues for this stable solution are of the form: $\theta_{\alpha} = \lambda - r_{\alpha}<0$, confirming the change of stability between solutions.

Alternative fixed points that correspond to any other $\alpha \neq k$ less-quality winning site have, at least, $k - \alpha$ positive eigenvalues, indicating that these solutions are actually saddle points. For instance, the eigenvalues for the winning option $\alpha = k-1$ are:
\begin{eqnarray*}
    & & \theta_m = r_{k-1} - r_m < 0 \quad m = 1,...,k-2 \\ 
    & & \theta_{k-1} = r_{k-1} - \lambda \\ 
    & & \theta_k = r_{k-1} - r_k > 0
\end{eqnarray*}

\subsection{Imperfect quality assessment: The case of non-null parameter \texorpdfstring{$\mu$}{mu}}\label{app:mu}
As we explained in the main text, the parameter $\mu$ in the orginal LES model governs the bees' quality assessment behavior. When $\mu \rightarrow 0$, bees independently assess the quality of sites, and thus, the duration of the advertisement period strongly correlates with the actual site quality. Conversely, when $\mu \rightarrow 1$, bees do not individually assess site qualities but rather imitate their peers without regard to the options' qualities. Consequently, the advertisement duration is influenced by a generic parameter $K$ that is the same for all sites, as indicated in Eq.~\eqref{eq:uncomm_rate}. Although we have primarily considered $\mu = 0$ in the text to explore the interplay between other parameters such as interdependence, discovery probabilities, and qualities, here we provide a brief overview of how the results change when nonzero values of the independent quality assessment parameter, $\mu$, are considered. For simplicity and in accordance with the prescription by List et al. in the original formulation of the model \cite{list_independence_2008}, we set $K$ equal to the maximum quality among all sites, i.e., $K = q_k$.

The deterministic solutions described in Section~\ref{sec:analytical_sol} are expressed as functions of the rates, $r_\alpha$, and therefore remain valid for any value of $\mu$. However, the results presented in the subsequent analysis of the model, particularly in the binary decision problem concerning consensus and stationary time, were obtained with $\mu = 0$. After introducing $\mu \neq 0$, achieving the same levels of consensus would simply require increased values of interdependence compared to the case of $\mu = 0$. Figure~\ref{fig:Qlines_var_mu} illustrates how the consensus crossover lines change, shifting towards greater $\lambda^{*}$, when varying $\mu$ for both symmetric and asymmetric discovery scenarios.

The condition to achieve consensus without interdependence, i.e. by simply relying on the different advertisement times for each option, $q_1 < q_2/x$, is now modified as follows. For the symmetric discovery scenario, either fixing $(q_2, \mu)$ or $(q_1,q_2)$, we obtain:

\begin{equation}
     \lambda = 0\text{:} \qquad q_1 \leq q_2 \frac{1-\mu}{x - \mu} \quad \text{or} \quad \mu \leq \frac{q_2 - 2 q_1}{q_2 - q_1} \text{.}
\end{equation}

For the asymmetric discovery probabilities, fixing $(\pi_1, q_1, q_2, \mu)$ or $(\pi_1, \pi_2, q_1, q_2)$, respectively, we obtain:
\begin{equation}
    \lambda = 0\text{:} \quad \pi_2 \geq \frac{\pi_1 x}{\mu + \frac{q_2}{q_1}(1-\mu)} \quad \text{or} \quad \mu \leq \frac{q_2 - q_1 \frac{\pi_1}{\pi_2}x}{q_2 - q_1}\text{.}
\end{equation}

\subsection{Criterion for measuring convergence times to the stationary state}\label{sec:appendixD}

The criterion for measuring the time when the system reaches the stationary state is as follows: We monitor the time evolution of each population fraction $f_{\alpha}(t)$, including the uncommitted state, and calculate the average of $f_{\alpha}$ over a time window of size $w$, i.e., over the time interval $[t, t+w]$. We consider a population to have reached the stationary state if the absolute difference between this average value and the value of $f_{\alpha}$ at the subsequent time step is below a certain threshold $\kappa$. In other words, if:

\begin{equation}
    | f_2(t+w+1) - \langle f_2 \rangle_{t,w+t} | < \kappa,
\end{equation}
we identify the stationary time for this particular population as $t_{ss} = t+w+1$. The longest time among all populations will be considered the effective simulation's stationary time. For our simulations in the binary decision problem, we have used a block size of $w=50$ and a threshold $\kappa = 5 \times 10^{-4}$. Other values within the intervals $w \in [25,100]$ or $\kappa \in [5 \times 10^{-4}, 5 \times 10^{-3}]$ produce qualitatively similar results.

\subsection{Lattice simulations in \texorpdfstring{$d=2$}{d=2}}\label{sec:appendixE}

\begin{figure}[t]
    \centering
    \includegraphics[width=.7\columnwidth]{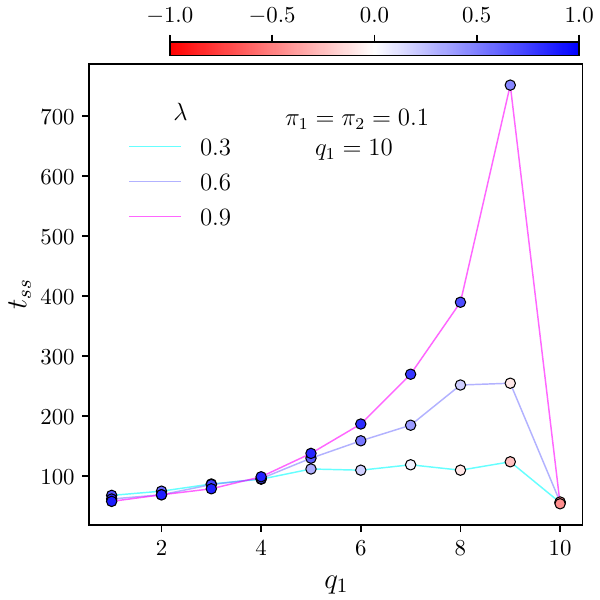}
    \caption{Time to reach the stationary state, $t_{ss}$, plotted as a function of the bad-quality option, $q_1$, while maintaining $q_2 = 10$ in lattice simulations. Lattice size is  $L=1024$. As in the main text, three values of the interdependence parameter have bee considered, namely $\lambda = 0.3, 0.6, 0.9$, with discovery probabilities $\pi_1 = \pi_2 = 0.1$.}
    \label{fig:lattice_times}
\end{figure}

We additionally implement a spatially distributed version of the model described in Sec~\ref{sec:modelling}, as a cellular automaton.
In this spatially distributed version of the model, each scout 
bee occupies a specific location on a square lattice with periodic boundary conditions. 
The lattice has a lateral length of $L$, resulting in a total of $N=L\times L$ nodes.
Each scout bee interacts only with its nearest neighbors on the lattice.

The system is defined by a scalar field $\{s\}$ where 
$s_i$ ($i=1,\ldots,N$) stands for the state of each bee $i$
and corresponds to a specific location ${\bf r}=(x,y)$ on the lattice, 
denoted as $s_i(t) \equiv s({\bf r},t) \equiv s(x,y; t)$.
Starting from an initial configuration $\{s\}(t=0)$, the system evolves
according to the update rules defined in Eqs.~\eqref{eq:state_evolution} and 
~\eqref{eq:comm_prob}.
The fractions $f_{\alpha,t}$ of bees promoting site $\alpha$ at time $t$ 
entering in Eq.~\eqref{eq:comm_prob}, are now 
computed in the square lattice as:
\begin{equation}
    f_{\alpha}({\bf r},t) = \frac{1}{4} \sum_{<n.n. {\bf r'}>} \delta(\alpha, s({\bf r'},t)),
\end{equation}
where $\delta$ is the Kronecker's delta, the sum runs over the four
nearest neighbors of ${\bf r}$ on the square lattice, and the index $\alpha = 1, \ldots, k$ indicates the possible committed states. The state update rules at each site, as described by Eq.~\eqref{eq:state_evolution}, remain the same.
A time step consist in updating every local state $s_i$ probabilistically
(Eq.~\eqref{eq:state_evolution}).
As a remark, we computationally accelerate this update process by using a 
massively parallel approach and GPUs as a hardware.
The technical approach to ensuring the correctness of the parallel implementation involves using a checkerboard lattice decomposition. This method alternates between updating all {\it black} cells in parallel and all {\it white} cells in parallel, preventing the overwriting of state values before they are used to compute transition probabilities. This parallelization technique ensures consistency with a serial implementation, where each cell is updated one at a time, maintaining equivalence at the global step balance level. It has been proven effective in various simulations, such as Monte Carlo~\cite{FerreroCPC2012} and Ginzburg-Landau~\cite{KoltonPRB2023} simulations. Additionally, observables such as population fractions, the order parameter, and fluctuations are calculated using parallel reduction algorithms.

The spatially distributed model, similar to \cite{Galla_2010}, is not designed to replicate realistic spatial behavior of bees but rather to analyze the dynamic competition of states within a spatio-temporal framework. As discussed in the main text, the general stationary results of this spatial model align perfectly with mean field (i.e., fully-connected) simulations. Analyses such as the time to reach the stationary state, as depicted in
Fig.~\ref{fig:sym_Q_and_time}F or Fig.~\ref{fig:asym_pi_Q_lines_and_time}E, also yield qualitatively similar results, exhibiting the same trends discussed in the main text - see Fig.~\ref{fig:lattice_times}. 

On the other hand, two-dimensional lattice simulations enable the study of critical behavior and some critical exponents around the absorbing phase transition occurring in equal-quality, and almost fully-interdependent, scenarios. In this limit, lattice simulations provide valuable insights into the universality class of the phase transition and help classify the critical behavior of the system beyond the mean-field approximation, as these simulations incorporate spatial fluctuations.


%

\end{document}